%% file: main.tex
\documentclass[aps,amsmath,preprintnumbers,amssymb,twocolumn,superscriptaddress,longbibliography,nofootinbib,showpacs]{revtex4-1}

\include{preamble.tex}

\begin{document}

\title{Nuclear Recoil Identification in a Scientific Charge-Coupled Device}

\input{authors.tex}

\date{\today}

\begin{abstract}
Charge-coupled devices (CCDs) are a leading technology in direct searches for dark matter because of their eV-scale energy threshold and $\mathrm{\mu}$m-scale spatial resolution. Recent studies have also highlighted the potential for using CCDs to detect coherent elastic neutrino-nucleus scattering (CE$\nu$NS). The sensitivity of future CCD experiments could be enhanced by distinguishing nuclear recoil signals from electronic recoil backgrounds in the CCD silicon target. 
We present a technique for event-by-event identification of nuclear recoils based on the spatial correlation between the primary ionization event and the defect cluster left behind by the recoiling atom, later identified as a localized excess of leakage current under thermal stimulation.
By irradiating a CCD with an \ambe neutron source, we demonstrate $>93$\% identification efficiency for nuclear recoils with energies $>150$\,keV, where the coincident ionization events were confirmed to be nuclear recoils due to their topology.
The technique remains fully efficient down to 90\,keV, decreasing to 50\% at 8\,keV, and reaching ($6\pm2$)\% between 1.5 and 3.5\,keV.
Irradiation with a \iso{24}Na gamma-ray source does not result in any detectable defect clusters, with the fraction of electronic recoils with energies $<85$\,keV that are spatially correlated with defects $<0.1$\%.    
\end{abstract}

\maketitle

\section{Introduction}\label{sec:introduction}

The detection of low-energy interactions of weakly interacting particles with atomic nuclei provides a means to search for the particles that may constitute the universe’s dark matter~\cite{DMDD2022, *Essig:2022dfa} and to measure coherent elastic neutrino-nucleus scattering (CE$\nu$NS)~\cite{CEvNS2022}. The experiments developed for this purpose use an instrumented target to detect the signal of a recoiling atom over backgrounds from environmental radiation, which are mostly electronic recoils from radioactive decays in the target and from the interactions of external gamma rays.
Therefore, discrimination between nuclear and electronic recoils at low energies is a powerful technique of background suppression.
Various realizations of nuclear/electronic recoil discrimination have been demonstrated in several targets, including cryogenic calorimeters~\cite{EDELWEISS:2016boq,SuperCDMS:2014cds,*SuperCDMS:2017mbc, CRESST:2019jnq} and noble liquids~\cite{DarkSide:2018kuk,LUX-ZEPLIN:2022xrq}.

Silicon charge-coupled devices (CCDs) are some of the most sensitive ionization sensors~\cite{Holland:2003kiw, Tiffenberg:2017aac} but so far have lacked the capability to discriminate between nuclear and electronic recoils.
Nevertheless, the DAMIC detector\textemdash a CCD array operating in a low-background environment deep underground\textemdash performed a highly sensitive search for low-mass dark matter particles that was competitive because of the low energy threshold of the detector~\cite{DAMIC:2020cut}.
Detectors based on CCDs have been successfully deployed at a short baseline from nuclear reactors to search for CE$\nu$NS~\cite{CONNIE:2014qlq, *CONNIE:2021ggh}, although they have yet to reach the sensitivity required for a positive detection.
The potential of CCDs in the search of CE$\nu$NS at the European Spallation Neutron Source has also been noted~\cite{Baxter:2019mcx}.
In all these cases, electronic-recoil backgrounds remain a significant limitation for CCD experiments.

In this paper, we demonstrate for the first time event-by-event identification of nuclear recoils in a CCD by making use of the spatial correlation of the primary ionization event with the cluster of defects generated in the silicon lattice by the recoiling atom that is later identified by thermal stimulation. This work builds on previous studies of neutron interactions with silicon indicating the potential for using crystal defects as a method for detecting the nuclear recoils from dark matter interactions~\cite{Lee:2022sxx, *Steven-thesis}.
Since low-energy electronic recoils are not expected to generate clusters of defects, this strategy can effectively be employed for nuclear/electronic recoil discrimination in CCD experiments.

\section{Methodology}\label{sec:methodology}

Charge-coupled devices are pixelated sensors with a fully depleted active silicon volume.
Free charges generated in the active volume by ionizing particles are drifted by the electric field and collected on the pixel array.
Since charges diffuse laterally as they drift, energy depositions that occur deeper into the CCD volume result in more diffuse patterns of charge on the pixel array.
After a user-defined exposure time, the pixel array is read out to generate an image, where each pixel value above the image pedestal is proportional to the charge collected by the pixel during the exposure.
The images are analyzed to identify clusters of pixels with charge.
Low-energy recoils, for which the track length is much shorter than the pixel size, result in two-dimensional Gaussian clusters, whose integral is proportional to the energy $E$ of the event, whose spread $\sigma_{xy}$ is positively correlated with the depth ($z$) of the interaction, and whose mean corresponds to the $(x,y)$ coordinates of the interaction.
To minimize noise from leakage current across the biased device, CCDs are typically operated at low temperatures (from 100\,K to 150\,K) when recording ionization events.

In addition to the primary ionization event, a nuclear recoil induced by a neutron (or weakly interacting particle) will produce a cluster of crystal defects in the silicon lattice by dislocating atoms along its path until it stops, with nuclear recoils of 2\,keV already producing clusters of up to 30 defects~\cite{Nordlund1998, Moll:1999kv, Srour2003, Sassi:2022njl}.
Conversely, electronic recoils must have at least $\sim$260\,keV of energy to dislocate single atoms and produce point defects, and at least $\sim$8\,MeV to produce defect clusters~\cite{Moll:1999kv, Pintilie:2002ge, Fretwurst:2002gd}.
For clarity, we refer to the clusters of defects that we detect simply as ``defects'' to distinguish them from ``clusters,'' which refer to contiguous pixels with charge observed in a CCD image. 
Such defects are small relative to the CCD pixel size and can persist in the silicon after the disordered state of the lattice stabilizes.
Defects in the silicon lattice distort the local band gap structure, resulting in midband energy states that give rise to excess leakage current~\cite{Srour2003}, which increases rapidly with temperature and can result in visible clusters above the shot noise at sufficiently high temperatures (e.g., 220\,K).
As is similar for ionization events, defects are measured as two-dimensional Gaussian clusters, whose integral is the total charge from the leakage current integrated over the exposure time.

In this study, we first acquired images with the CCD at warm temperatures (221\,K) to identify existing defects. We then lowered the temperature of the CCD and proceeded with a series of cold images acquired while the CCD was irradiated with an \ambe neutron source to measure the ionization signals from nuclear recoils. Finally, the temperature was increased back to the original value for a second series of warm images to identify the defects generated during the irradiation. The data were analyzed to search for correlations in the $(x,y)$ coordinates between ionization clusters in the images during irradiation and clusters from defects that appeared following the irradiation. The experiment was repeated with a \iso{24}Na gamma-ray source to characterize backgrounds due to electronic recoils and confirm that these recoils do not generate visible defects, and a third time without any source to characterize the effect of environmental backgrounds. 

In Section~\ref{sec:setup}, we provide the details of the experimental setup, the warm and cold data sets, and the images for analysis.
In Section~\ref{sec:recspec}, we describe how we reconstruct the energy spectrum of nuclear recoils induced in the CCD by neutrons from the \ambe source. We first reconstruct the high-energy part of the spectrum by selecting nuclear-recoil clusters based on their topology (Sec.~\ref{sec:topology}) and then extrapolate toward lower energies by subtracting the expected electronic-recoil backgrounds from the \ambe source (Sec.~\ref{sec:lowespec}).
In Section~\ref{sec:defects}, we describe how we use the warm images to identify the defects generated during irradiation.
In Section~\ref{sec:coincidence}, we present the results from the search for spatial correlations between clusters from ionization events in the cold data and clusters from defects in the warm data, where we demonstrate that only nuclear-recoil ionization events from the \ambe source show a statistically significant correlation with defects.
Finally, in Section~\ref{sec:results}, we divide the measured spectrum of ionization events from the \ambe source that are spatially correlated with defects by the reconstructed total spectrum of nuclear recoils to obtain the fraction of nuclear recoils that generate visible defects as a function of energy.

\section{Experimental Setup and Data}\label{sec:setup}

\begin{figure}[t]    
    \includegraphics[width=1.0\linewidth]{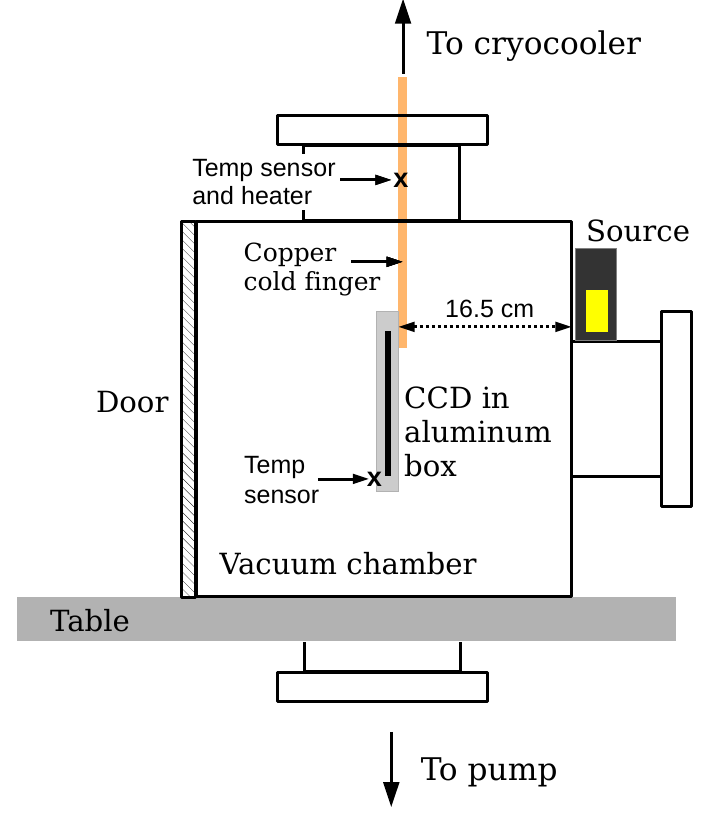}
    \caption{Cross section sketch of the experimental setup with \ambe source positioned outside the vacuum chamber and enclosed in a 6mm-thick lead vial.}
    \label{fig:chamber}
\end{figure}

All data were acquired in a surface laboratory on the University of Washington campus in Seattle. The 24-megapixel CCD ($6144\times 4128$ pixels, $15\times 15$ $\mu$m$^{2}$ pixel size, 670 $\mu$m thick) was developed by Lawrence Berkeley National Laboratory MicroSystems Lab~\cite{Holland:2003kiw} for the DAMIC-M dark matter direct detection experiment \cite{DAMIC-M:2023gxo}.
The CCD system is housed in a stainless steel vacuum chamber that is evacuated to $\simeq 10^{-5}$ millibar. 
Inside, the CCD is kept fixed in an aluminum storage box, which is screwed onto a copper cold finger attached to a Cryotel GT cryocooler.
A temperature sensor and heater on the cold finger are connected to an external PID controller that maintains the temperature at a set point.
A second sensor monitors the temperature of the storage box, which is estimated to be $<$5\,K lower than the CCD silicon temperature from thermal simulations.
The system was operated at a storage-box temperature in the range 147\,K to 221\,K.
Electrical cables carry the signals to/from the outside electronics through a vacuum feedthrough.
A 111 MBq \ambe neutron source (mean neutron energy 4.2\,MeV; neutron rate 7400\,s$^{-1}$) was used to generate nuclear recoils in the bulk silicon of the CCD. To attenuate the flux of gamma rays from the \ambe source and to allow for easy removal, the source was enclosed in a lead vial with wall thickness of 6\,mm and positioned outside the vacuum chamber, as shown in Fig.~\ref{fig:chamber}.

Lateral charge diffusion limits the sensitivity to low-energy recoils and defects because it distributes the charge over multiple pixels.
Since \sxy\ is inversely proportional to the square-root of the substrate bias~\cite{Holland:2003kiw}, we operated the CCD at the maximum bias of 100\,V.

The CCD was read out by clocking charge row-wise into the horizontal register, where the charge was clocked pixel by pixel to two charge-to-voltage amplifiers located at opposite ends of the horizontal register for charge measurement.
The CCD can be operated in skipper mode, whereby multiple non-destructive charge measurements (NDCMs) of a single pixel are performed, suppressing the readout noise averaged over $N$ measurements, $\sigma_{r}$, by $1/\sqrt{N}$.
We used a commercial CCD controller from Astronomical Research Cameras, Inc., to supply the clocks and biases and to measure the pixel values with a noise of $\sigma_{r} \approx$ 6\,$e^-$ for $N$ = 1.

\subsection{Data Sets}
\label{sec:dsets}

Images were acquired in dedicated data sets at two different temperatures, the details of which are given in Table~\ref{tab:sources}.
Each CCD image was exposed for 20 minutes, during which time the clocks were stopped to allow charge to accumulate on the pixel array, followed by readout.
First, a series of 31 warm ``pre-irradiation" images were taken at $221\pm3$\,K with a single measurement per pixel, $N=1$, resulting in a readout time of 4 minutes.
The CCD was then cooled at a rate of 0.3 K/min to $147\pm1$\,K, and 10 cold images were acquired with the \ambe source in place.
To reduce readout noise, the CCD was operated in skipper mode, with $N=10$, resulting in $\sigma_{r} \sim$ 2$e^{-}$ and a readout time of 28.7 minutes per image.

To preserve the spatial correlation between nuclear recoil events as they appear in an image and their physical location on the CCD, the \ambe source was removed and shielded during each readout.
The CCD was then warmed to $221\pm3$\,K at a rate of 0.5 K/min. 
Leakage current in the CCD scales exponentially with temperature~\cite{janesick2001scientific} and is a more sensitive probe for the CCD temperature than the temperature sensor on the storage box.
Thus, the temperature was manually adjusted about the nominal value of 221K until the leakage current was consistent with the pre-irradiation images, and a second set of 31 ``post-irradiation" images was acquired.
The experiment was repeated a second time with a 3.7\,kBq \iso{24}Na gamma-ray source (energies 1.37\,MeV, 2.75\,MeV) in place of the $^{241}$Am$^{9}$Be, and a third time with no source to characterize environmental backgrounds.
Each warm data set took approximately 13 hours to acquire, and the temperature stability was within 0.3\,K for the \ambe data sets and 0.7\,K for the \iso{24}{Na} and background data sets.

The calibration constant to convert raw pixel values to number of electrons was obtained for each amplifier from an image read out with $N=500$, where the readout noise $\sigma_{r}$ = 0.23\,e$^{-}$ was sufficiently low to identify discrete peaks for the charge in the pixels~\cite{DAMIC-M:2022xtp}. 
This measurement was performed at a CCD temperature of $147$\,K prior to acquiring each of the three sets of cold images, and the calibration constant was found to be stable within 2$\%$.

\begin{table}[t]
    \begin{center} Cold Data \\ \end{center}
    {\begin{tabular}{cccc}
    
      \hline\hline
     ID & No. images & \spix\ & \\
        &        &  [e$^{-}$] &\ \\
     \hline 
     \ambe & 10  &  1.8$\pm$2&\\
     \iso{24}Na & 10  & 1.6$\pm$1&\\ 
     bkgd & 10 &  1.6$\pm$1&\\
     \hline \hline
     \end{tabular} }
     \begin{center} Warm Data \\ \end{center} 
    {\begin{tabular}{cccc}
     \hline \hline
     ID & No. images & \spix\ & Leakage charge\\
        &      &         [e$^{-}$] & [e$^{-}$]      \\
     \hline 
     pre-\ambe & 31  & 198$\pm$13 & 1690$\pm$290  \\ 
     post-\ambe & 31  & 192$\pm$11 & 1672$\pm$295 \\
     pre-\iso{24}Na &  31 & 140$\pm$11 & 1621$\pm$181\\ 
     post-\iso{24}Na & 31  & 141$\pm$9 & 1710$\pm$195\\ 
     pre-bkgd & 31  & 139$\pm$6& 1686$\pm$186\\ 
     post-bkgd & 31  & 144$\pm$3& 1718$\pm$193 \\ 
     \hline \hline

    \end{tabular}}
    \caption{Summary of images taken at $147\pm1$K (top) and $221\pm3$K (bottom).
    Cold 147\,K images were acquired during \ambe or \iso{24}Na irradiation to measure the primary ionization events with $N$ = 10 NDCMs per pixel.
    Warm 221\,K images were acquired pre- and post-irradiation with $N$ = 1.
    The background (bkgd) images were acquired in the same manner but without a source.
    The number of images in the data set, leakage charge, and pixel noise (\spix ) in the images are provided.
    }
    \label{tab:sources}
\end{table}

\subsection{Images and Masks}
\label{sec:imasks}

All images consist of 6400 columns and 2000 rows, with each amplifier reading 3200 pixels in the horizontal direction, including 128 past the end of the physical pixel array.
These 128 columns are referred to as the ``overscan'' and correspond to measurements of empty pixels that contain only readout noise.
During image readout the CCD continues to collect ionization charge from environmental backgrounds, which results in a higher density of background events in the region of the image that is read out last.
Thus, we only read 2000 out of the 4128 physical rows of the CCD to decrease the number of pileup events. 
To remove the charge left over after this partial readout, the full CCD pixel array was cleared of charge before beginning a new image by rapidly clocking the charge toward the amplifiers and dumping it without measurement.
The background leakage charge, which is the average number of electrons per pixel accumulated from the leakage current during exposure, is measured by taking the difference between the average pixel value in a background region of the pixel array with no ionization events or defects and the overscan.
The pixel noise in the images, \spix, was estimated from the standard deviation of the pixels in the background region.
The pixel noise has contributions from both readout noise, $\sigma_{r}$, and statistical fluctuations from background leakage charge (shot noise), which is dominant in the data acquired at higher temperatures and does not decrease with increasing number of NDCMs.

For each of the six warm data sets in Table~\ref{tab:sources}, we generated a ``median image,'' where each pixel value is the median of the given pixel over all images in the data set.
Defects, which appear at the same location across the data set, are most readily identified in the median images, while ionization events, which appear only in a single image, are effectively filtered out.
Pre-existing defects can originate during fabrication or may arise over the lifetime of the CCD.
They may be stable over time or may disappear after temperature-cycling (annealing) the CCD to room temperature. Some prominent, stable defects have enough charge to overcome potential barriers when the charge is shifted during readout, causing vertical streaks (``hot columns"),
which can interfere with cluster identification.
To exclude image regions affected by pre-existing defects, we generated a list of pixels, referred to as a ``mask," from the median images, with a separate mask for each of the three experiments.
The masks include regions where the pixel value exceeds by 3\,\spix\ the average value of background pixels in both pre- and post-irradiation median images. Pixels on the edges of the image with coordinates $x\le$10 or $x\ge6391$ and $y\le10$ or $y\ge1991$ were also masked to exclude noise and baseline transients at the beginning of image readout and after row shifts.

\section{Nuclear Recoil Ionization Spectrum}\label{sec:recspec}

Cold images were processed by first averaging over all NDCMs of every pixel and then subtracting the pedestal, representing the average analog-to-digital (ADC) value of the background noise pixels.
The pedestal was calculated separately for each column segment of 1000 consecutive pixels by fitting to a Gaussian function the lowest, most prominent peak in the pixel-value distribution.
The mean value from the fit was then subtracted from each pixel in the column segment and the process was repeated for row segments of 800 consecutive pixels.

Ionization events in the cold images may be produced by neutrons and gamma rays from the radioactive sources and from environmental radiation.
We identify ionization events in the cold (147\,K) images as contiguous pixels each with value $>4$\,\spix. We exclude from the analysis any cluster directly adjacent to a masked pixel. 
There is an upper limit on the size of a charge packet that can be efficiently transferred to the readout stage and its value measured repeatedly without charge loss.
By comparing the first and second out of 10 NDCMs for each pixel in an image and noting where the difference exceeded the readout noise, we determined saturation to occur at 5530$\pm$80 e$^{-}$ for one amplifier and 5250$\pm$125 e$^{-}$ for the other.
Since saturation affects the reconstruction of the energy and topology of the cluster, any cluster containing at least one pixel above 5070\,e$^{-}$ was omitted from the analysis. 
The total charge of every cluster was estimated by summing over the pixel values.
We also evaluated the charge-weighted mean and variance of the pixel coordinates to obtain the cluster $(x,y)$ location in the image and the \sxy\ spread of the cluster, respectively.

The total charge of the event was converted to deposited energy $E$ (in ``electron-equivalent'' units \eve ) by considering that an electronic recoil ionizes on average one electron-hole pair for every 3.8\,eV of energy deposited~\cite{Ramanathan:2020fwm}. The corresponding nuclear-recoil energy in \evr\ was obtained from the electron-equivalent values using the parameterized model from Ref.~\cite{Sarkis:2022pvc}.
To determine our sensitivity to low-energy events in our data, we simulated point-like events distributed uniformly in the CCD volume and introduced them on top of noise-only ``blank'' images.
To relate the depth of the interaction with the simulated \sxy , we used the diffusion model outlined in Ref.~\cite{DAMIC06KGD}, with parameters obtained from muon tracks acquired with a similar 24-megapixel CCD and scaled to our substrate bias of 100\,V.
By running our clustering algorithm on the simulated images, we obtain a clustering efficiency $>$99$\%$ and accurate energy reconstruction down to 0.2\,k\eve . 

\subsection{Identification by Topology}\label{sec:topology}

Atoms recoiling after scattering with neutrons from the \ambe source have track lengths smaller than the pixel size, and the ionization events can be considered to be point-like, while recoiling electrons are only point-like below $\sim$85\,keV$\mathrm{_{ee}}$.
The extended tracks of electronic recoils above this energy can be easily distinguished from nuclear recoils by cluster topology alone, thereby allowing us to select a clean sample of nuclear recoils to construct the high-energy spectrum.

\begin{figure}[t]
    \includegraphics[width=\linewidth]{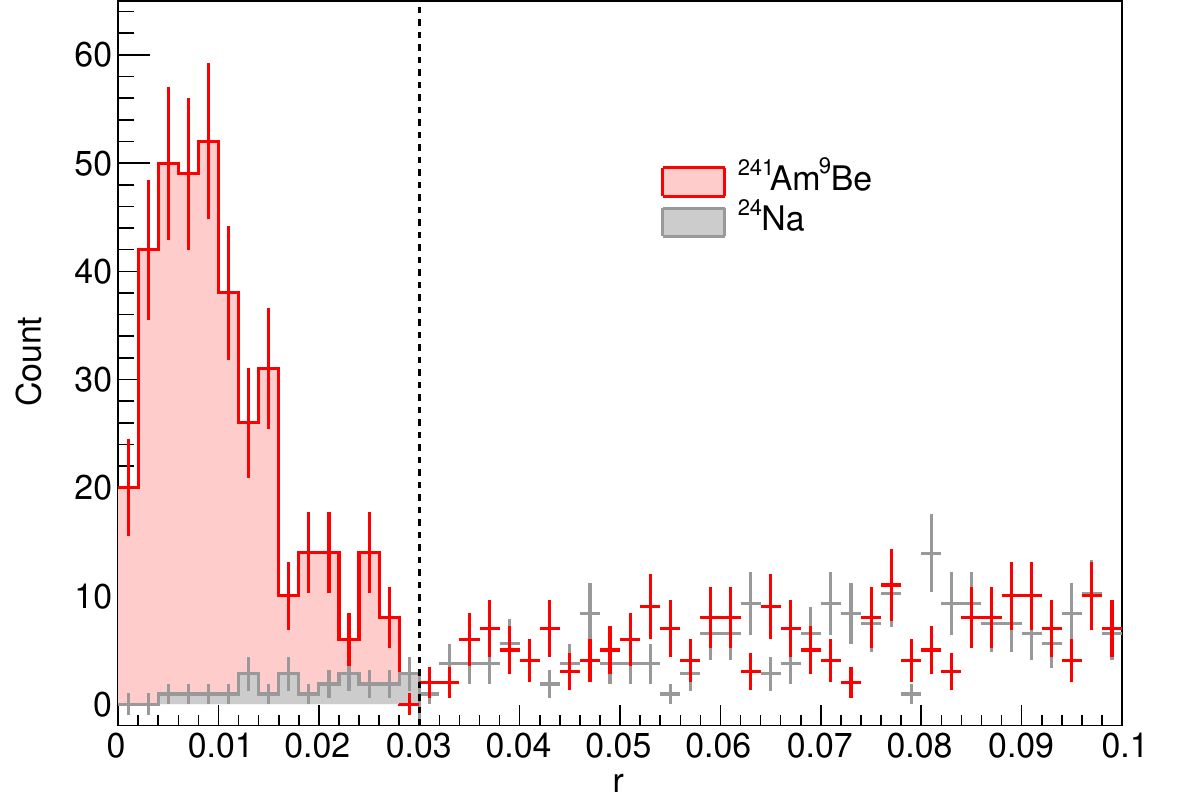}
    \caption{Cluster $r$ values (defined in the text) for \ambe (red) and \iso{24}Na (gray) ionization events above 85\,keV$\mathrm{_{ee}}$. Clusters with $r <$ 0.03 are identified as high-energy nuclear recoils. The integral of the gray filled histogram is 6$\%$ that of the red filled histogram and represents the estimated electronic recoils in the \ambe data after the selection.} 
    \label{fig:topo}
\end{figure}

The characteristic symmetry of nuclear recoil clusters can be parameterized by the ratio of the spread of charge in the vertical and horizontal directions, $\sigma_y/\sigma_x$, which is expected to be unity, and the Pearson correlation coefficient between the $x$ and $y$ coordinates of the pixels, $c$, which is expected to be zero.
The two variables are complementary since $\sigma_y/\sigma_x$ best identifies clusters that are preferentially along the horizontal or vertical directions, while $c$ best identifies clusters that are preferentially along a diagonal.

We define a single selection parameter $r = \sqrt{(1-\sigma_y/\sigma_x)^{2} + c^2}$ and classify events as high-energy nuclear recoils if $r<0.03$ for $E$ $>$ 85\,k\eve\ (150\,k\evr). This selection was chosen by comparing clusters in the \ambe and background-only data sets and considering a region in $\sigma_y/\sigma_x$\textendash $c$ space containing a statistical excess of events in the \ambe data.

Figure~\ref{fig:topo} shows the distribution of $r$ for clusters with $E>85$\,k\eve\ in the \ambe and \iso{24}Na data, where the \iso{24}Na histogram was scaled such that its integral for $0.03 < r < 0.1$ matches the integral of the \ambe histogram in this region. 
We find 374 clusters with $r<0.03$ in the \ambe data.
Comparing the \ambe and scaled \iso{24}Na histograms, 
we conclude that $(6.0\pm 1.2)$\% of the selected clusters are electronic recoils.
The dark blue histogram in Fig.~\ref{fig:spec} shows the high-energy spectrum of nuclear recoils from the \ambe source identified by cluster topology, where the estimated leakage from electronic recoils, approximated with the spectrum of clusters in the \iso{24}Na data, was subtracted.

\subsection{Extrapolation to Low Energies}\label{sec:lowespec}

To determine the nuclear-recoil spectrum below 85\,k\eve , we subtract the contribution of electronic recoils from the \ambe source and from environmental background from the spectrum of all clusters in the \ambe data.
Electronic recoils from the \ambe source are dominated by the primary 4438\,keV gamma rays emitted by the deexcitation of \iso{12}C\,$^*$ following $\sim$58\% of $(\alpha,n)$ reactions~\cite{liu:2007}.
The prominent 59.5\,keV gamma rays from $^{241}$Am decay are fully attenuated by the lead vial, while higher-energy gamma rays have negligible intensities~\cite{Eriksen:2020isg, BASUNIA20062323}.
Secondary gamma rays from the inelastic scattering of fast neutrons from the source and from the capture of thermal neutrons with nuclei in the setup may also produce electronic recoils.

\begin{figure}[t]
   \includegraphics[width=\linewidth]{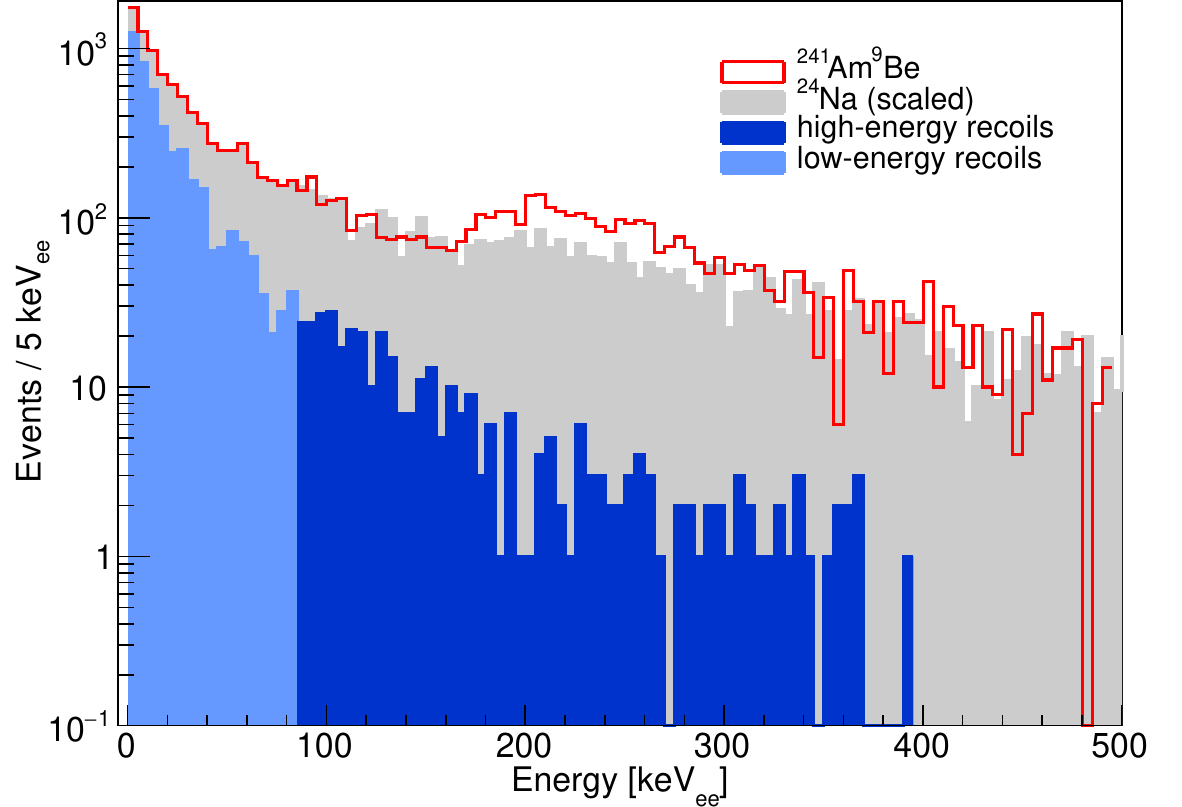}
   \caption{Total background-subtracted \ambe spectrum (red) and reconstructed spectrum, consisting of the scaled, background-subtracted \iso{24}Na electronic-recoil spectrum (gray) stacked atop the nuclear-recoil spectrum (blue).
   Above 85\,k\eve\ (dark blue), the nuclear recoils are the events identified by topology.
   Below 85\,k\eve\ (light blue), we assume the nuclear recoils to be the difference between the \ambe and scaled \iso{24}Na spectra.}
   \label{fig:spec}
\end{figure}

\begin{figure}[t]
   \includegraphics[width=\linewidth]{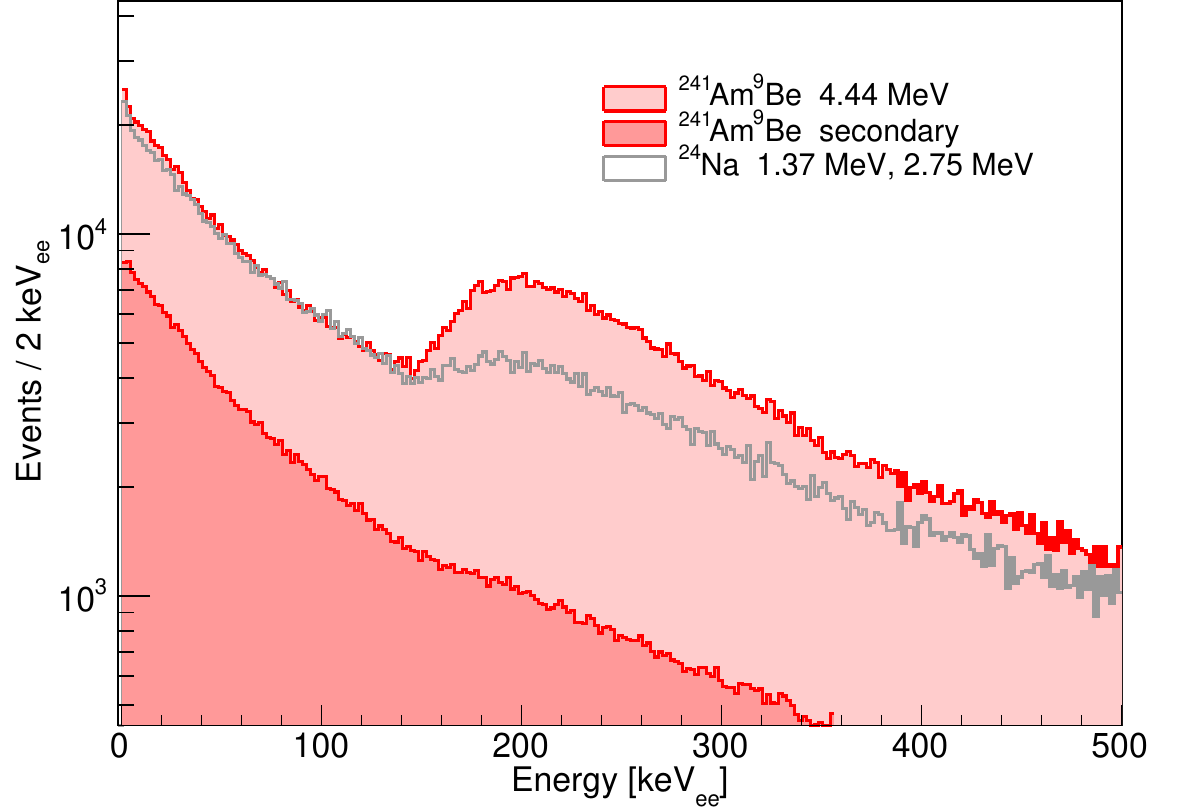}
   \caption{Simulated spectrum of electronic recoils in the CCD from gamma rays from the \ambe (stacked red histogram) and \iso{24}Na (gray) radioactive sources.
   The spectrum from \iso{24}Na was scaled to match the integral of the \ambe spectrum between 85 and 140\,k\eve .}
   \label{fig:gcomp}
\end{figure}

To estimate the contribution from gamma rays to the spectrum, we simulated with {\tt Geant4}~\cite{GEANT4:2002zbu} primary neutrons and gamma rays radiated by the \ambe\ and \iso{24}Na sources and propagated them through a model of our experimental setup, including the detailed geometry and material composition.
The neutron and gamma-ray spectra for the \ambe and \iso{24}Na sources were obtained from Ref.~\cite{AmBeSpec,*Duke:2015wga} and~\cite{Basunia:2022zmp}, respectively.
Within {\tt Geant4} version 10.04, the Livermore low-energy electromagnetic models were used to control the electron and gamma-ray transport and interactions.
The low-energy Neutron High Precision (HP) package was used for neutron transport, scattering, and capture.
Our {\tt Geant4} simulation shows that electronic recoils from secondary gamma rays, mostly from inelastic scattering in the stainless steel vacuum chamber (74\%), cold finger (9\%), and lead vial (9\%), contribute 26\% of all electronic recoils from the \ambe\ source.
Figure~\ref{fig:gcomp} shows the electronic-recoil spectrum from primary and secondary gamma rays from the \ambe\ source, which is very similar to the spectrum from \iso{24}Na below 140\,k\eve , with a maximum difference of 10\%.
Therefore, we use the measured spectrum from \iso{24}Na as a model of the electronic-recoil background from the \ambe source below 140\,k\eve\ and consider the systematic uncertainty from this choice in Sec.~\ref{sec:results}.
Using the \iso{24}Na data spectrum accounts for detector effects (e.g., noise, pixel saturation, cluster reconstruction, cluster selection, etc.) and inaccuracies in the simulation that equally affect gamma rays from the \ambe\ and \iso{24}Na radioactive sources.

Figure~\ref{fig:spec} shows the measured ionization spectrum from the \ambe source and the reconstructed spectrum obtained by adding the electronic-recoil spectrum from the \iso{24}Na source to the nuclear-recoil spectrum.
The environmental background was subtracted from both the \ambe and \iso{24}Na spectra.
The \iso{24}Na spectrum was scaled in amplitude so that the addition of the \iso{24}Na spectrum to the high-energy nuclear recoils identified by topology matches the total \ambe spectrum in the range 85\,k\eve $< E <$ 140\,k\eve .
The difference between the scaled $^{24}$Na spectrum and the total $^{241}$Am$^{9}$Be spectrum is then the spectrum of nuclear recoils below 85\,k\eve\ down to our 0.2\,k\eve\ threshold (light blue histogram in Fig.~\ref{fig:spec}).

\section{Defect Identification}\label{sec:defects}

As discussed in Sec.~\ref{sec:imasks}, defects in the CCD are most readily identified in the median images.
To identify defects that appear during the irradiation of the CCD, we generated a ``difference image'' where each pixel is the difference between the warm post- and pre-irradiation median images 
in each of the three experiments.
The noise \spix\ in the median difference images, presented in Table~\ref{tab:coinc}, 
is much lower than in the individual warm images (Table~\ref{tab:sources}), which results in significant improvement in the sensitivity to defects.

\begin{figure}[t]    
\includegraphics[width=\linewidth]{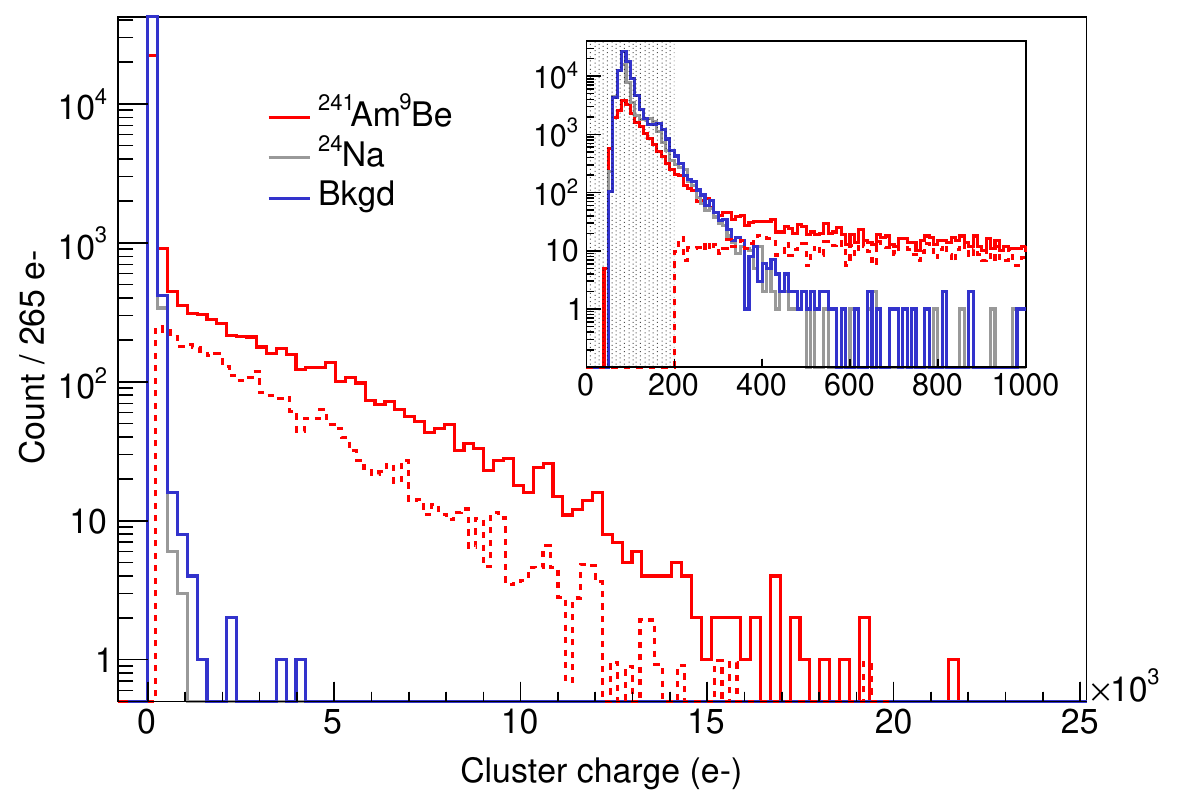}
    \caption{Charge of clusters identified in the median difference image (Sec.~\ref{sec:defects}) of the warm \ambe (red), \iso{24}Na (gray), and background (blue) data sets.
    The long tail of high-charge clusters in the \ambe data corresponds to nuclear-recoil generated defects.
    The inset focuses on the region below 1000\,$e^-$, where the prominent peak below 200\,$e^-$ is clustered noise.
    We consider clusters with $>200$\,$e^-$ as candidate defects and exclude those in the shaded region of the inset.
    The dotted red lines show the defects that coincide with ionization events, minus accidentals, in the \ambe data (Sec.~\ref{sec:coincidence}).
    }
    \label{fig:defects}
\end{figure}

Candidate defects in the median difference image were identified by running a clustering algorithm that groups adjacent pixels with charge $>2.5$\,\spix\  if at least one pixel has charge $>$80 e$^{-}$ ($\sim$4\spix).
The total charge of the candidate was evaluated by summing the charge of the pixels in the cluster, while the cluster position was estimated as the charge-weighted mean of the $(x,y)$ coordinates of the pixels.
Clusters directly adjacent to a masked pixel were omitted from the analysis.
Figure~\ref{fig:defects} shows the charge distribution of clusters in the median difference images.
The \ambe\ spectrum shows a clear excess of clusters compared to the other experiments above $\sim$300\,$e^-$, which provides clear evidence of nuclear-recoil generated defects.
The inset shows the spectrum below 1000\,$e^-$, which is dominated by clustered noise just above the 80\,$e^-$ clustering threshold and decreases exponentially with increasing cluster charge.
We consider candidate defects to be all clusters with charge $>$\,200~e$^{-}$.
While lowering this value would increase the acceptance for defects, it would also increase the acceptance for noise clusters, which interfere with the spatial coincidence search (Sec.~\ref{sec:coincidence}).
The selection was chosen so less than 5\% of the spatial coincidences in the \ambe data are accidentals.
Table~\ref{tab:coinc} summarizes the number of candidate defects after $^{241}$Am$^{9}$Be irradiation, \iso{24}Na irradiation, and with no source (background).
Since noise clusters dominate the \iso{24}Na and background spectra, the fewer candidates in the \iso{24}Na data are because of the slightly lower noise in the median difference image, with no evidence of defect generation.
Conversely, there are $\sim$5000 visible defects above background that appear following \ambe irradiation.

\begin{table}[t]
    \centering
    {\begin{tabular}{c c ccc}
      \hline\hline
     Experiment & \spix &Defects & Coincidences & Accidentals \\
      &        [e$^{-}$] & & & \\
     \hline
     \ambe & 18$\pm$1 &6777 & 3580 & 168$\pm$13\\ 
     \iso{24}Na & 24$\pm$1& 1570 & 44& 40$\pm$6\\
     bkgd & 26$\pm$1&1879 & 22 &  28$\pm$5\\ 
     \hline \hline

    \end{tabular}}
    \caption{Pixel noise (\spix ) in each median difference image, total number of candidate defects, number of defects that coincide with a selected ionization event, and estimated accidentals for the three experiments.}
    \label{tab:coinc}
\end{table}

\section{Coincidence Search}\label{sec:coincidence}

To correlate ionization events with defects, we performed a simple coincidence search by comparing the $(x,y)$ coordinates of each ionization cluster selected in the cold data to that of all defect clusters above threshold in the corresponding warm data, requiring that the two locations fall within one pixel width apart.
To estimate the number of events that accidentally coincide with a defect, we performed the same coincidence search after replacing the coordinates of every defect with a random position in the unmasked region of the image.
Table~\ref{tab:coinc} shows the number of coincidences and the expected accidentals in each of the three experiments with the different source configurations.
Only the irradiation with the \ambe source shows a statistically significant number of spatial correlations, with an upper limit on the number of coincidences above accidentals in the \iso{24}Na data of $<19$ (95\% C.L.). 
Considering that there are $1.7\times10^{4}$ ionization events with $E<85$\,k\eve\ in the \iso{24}Na data, this corresponds to a fraction of point-like electronic recoils that are spatially correlated with a visible defect $<0.1\%$.
This is consistent with the expectation that electronic recoils below $\sim$260\,keV do not generate any defects (Sec.~\ref{sec:methodology}).
Thus, we conclude that the defects that arose during the \ambe irradiation were caused by nuclear recoils.

Of the $\sim$5000 defects above threshold that appear following \ambe\ irradiation, 3580 coincide with selected ionization events.
Another 969 coincide with an ionization event that contains at least one saturated pixel (909) or is directly adjacent to a masked pixel (60) and was already excluded from the analysis.
An additional 13$\%$ of defects ($\sim$650) do not coincide with a nuclear recoil because of pileup, \emph{i.e.}, the ionization event is clustered together with an overlapping event in the image and the mean location of the cluster is not the location of the nuclear recoil.
The effect of pileup was estimated by performing a coincidence search between a representative sample of nuclear recoils simulated on top of \ambe\ cold images and the simulated coordinates of the event as the location of a defect. 
To confirm that correlated events are not missed because the distance requirement between cluster centers is too small, we increased the distance to 2 pixels, which resulted in 3400$\pm$22 coincidences above accidentals, consistent with the result in Table~\ref{tab:coinc}.

Of the 3580 coinciding defects, 435 are coincident with a selected ionization event with $E>85$\,keV$_{\mathrm{ee}}$, with an estimated 71$\pm 8$ accidentals. Of these events, 338 were identified as nuclear recoils by cluster topology. We visually inspected the coincident high-energy ionization events that are not identified as nuclear recoils by topology and conclude that 14 are likely misidentified because of pileup with a low-energy event that distorts the cluster topology but does not significantly displace the mean position of the cluster. Another 5 resemble nuclear recoils that just barely fall outside our selection criteria, with a cluster $r$ value $<$ 0.04.
This leaves 78 coincidences that are not nuclear recoils based on topology, consistent with the 71$\pm 8$ accidentals.
Conversely, of the 374 total ionization events identified as nuclear recoils by topology, 36 of them do not coincide with a defect, consistent with the $22\pm 5$ electronic recoils that we expect to be misidentified as nuclear recoils.

Finally, we confirm that the coinciding ionization events are evenly distributed throughout the cold images.
The \ambe\ data were acquired continuously except for a four-hour break between the fourth and fifth (out of 10) images.
The number of coincidences above accidentals per image was $344 \pm 10$ for the first four and $339 \pm 8$ for the last six images, which suggests that defects remain stable for at least the 12 hours that the CCD temperature was kept at 147\,K.

\begin{figure}[t]
    \includegraphics[width=0.5\textwidth]{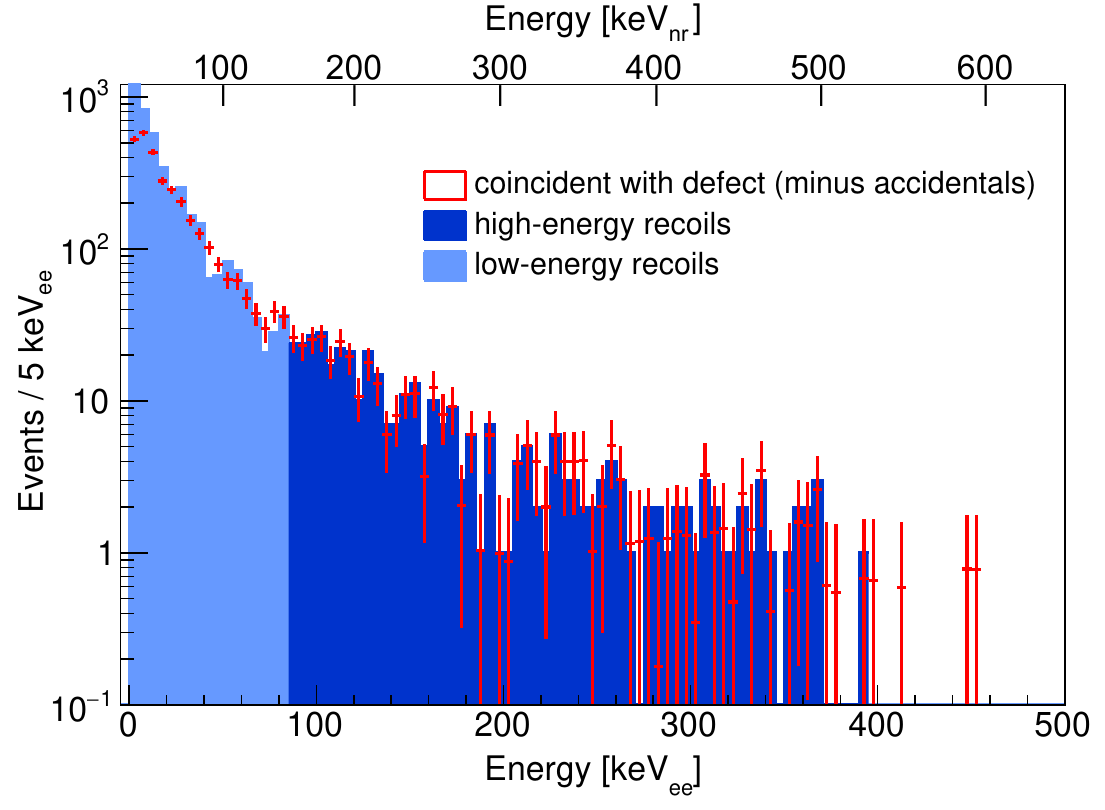}
    \caption{Reconstructed nuclear-recoil spectrum from the \ambe source (blue) compared with the spectrum of ionization events that are spatially correlated with defects (red).}
    \label{fig:defficiency1}
\end{figure}

The dashed red line in Fig.~\ref{fig:defects} shows the charge of defect clusters that coincide with ionization events minus accidentals. 
The difference between the red solid and dashed lines at high cluster charge are the defects that are missed because they either coincide with ionization clusters that have at least one saturated pixel or are missed altogether because of pileup.
Figure~\ref{fig:defficiency1} shows the spectrum of the corresponding ionization events (red markers), together with the low- (light blue) and high-energy (dark blue) nuclear-recoil spectra from Sec.~\ref{sec:recspec}.
In Fig.~\ref{fig:defficiency1}, the 19 coincident nuclear-recoil events that were misidentified by cluster topology have been added to the spectrum of high-energy nuclear recoils.
We consider the systematic uncertainty associated with this choice in Sec.~\ref{sec:results}.

\section{Nuclear Recoil Defect-Identification Efficiency}\label{sec:results}

\begin{figure}[t]
    \includegraphics[width=0.49\textwidth]{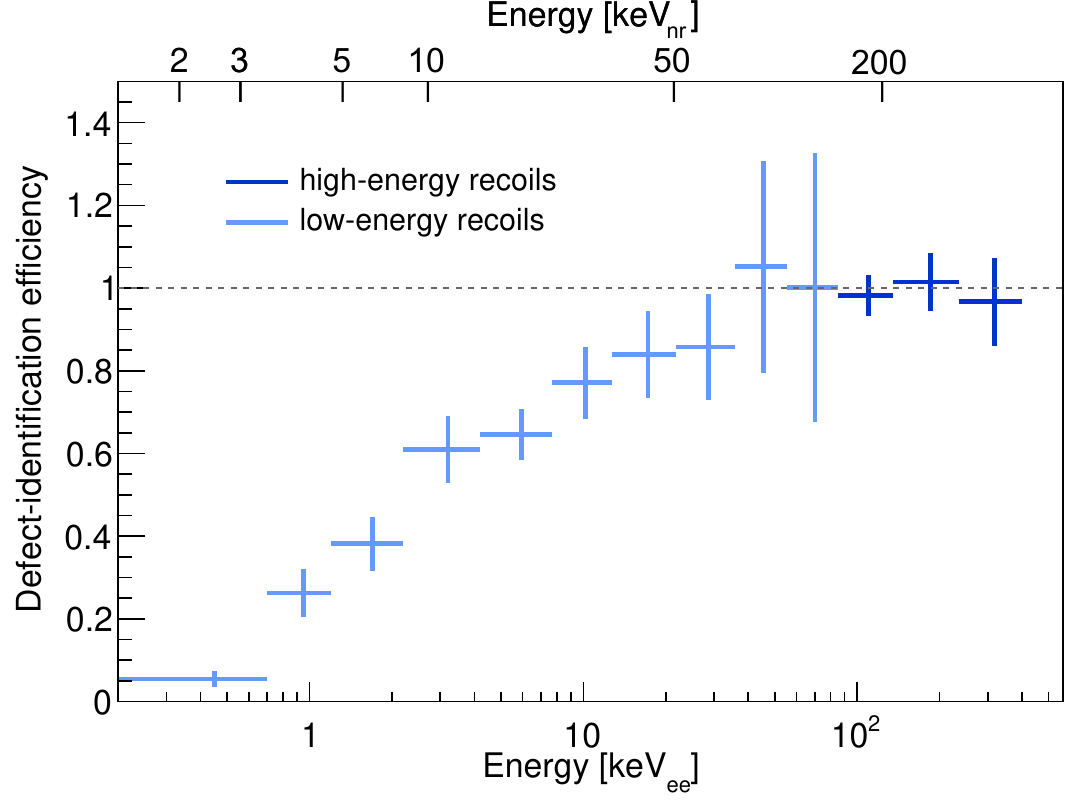}
    \caption{Fraction of nuclear-recoil ionization events that are spatially correlated with a defect above threshold as a function of energy.
    The corresponding fraction for electronic recoils is $<0.1\%$.
    }
    \label{fig:defficiency2}
\end{figure}

The ionization spectra of nuclear recoils and events that are spatially correlated with defects agree very well at high energies (Fig.~\ref{fig:defficiency1}), confirming full identification efficiency of nuclear recoils for $E>85$\,k\eve\ (150\,k\evr).
The agreement continues down to at least 45\,k\eve\ (90\,k\evr ), below which point there are fewer spatially correlated events.
To  obtain the efficiency in the identification of nuclear recoils from the spatial correlation between the primary ionization event and the defect, we divide the coincident spectrum by the nuclear-recoil spectrum.
Figure~\ref{fig:defficiency2} shows the resulting efficiency as a function of energy, with the electron-equivalent (nuclear-recoil) energy scale in the bottom (top) axis.
Above 85\,k\eve\ (150\,k\evr), $>93$\% (95\% C.L.) of nuclear recoils produce visible defects.
The efficiency starts decreasing at around 45\,k\eve\ (90\,k\evr ) to 50\% at 2\,k\eve\ (8\,k\evr ), and reaches ($6\pm2$)\% in the lowest-energy bin between 0.2\,k\eve\ (1.5\,k\evr) and 0.7\,k\eve\ (3.5\,k\evr).
This result can be compared to other detector technologies that feature nuclear/electronic recoil discrimination in the keV energy range~\cite{SuperCDMSSoudan:2013ukv, Strauss:2014zia, EDELWEISS:2017lvq, DarkSide:2018kuk, LUX:2020car, PICO:2022nyi}.

The systematic uncertainty in the defect-identification efficiency comes from the reconstruction of the nuclear recoil spectrum.
For $E>85$\,k\eve , the uncertainty arises from inefficiencies in the selection of nuclear recoils by topology, for which we attempted to correct by recovering  the 19 misidentified events in Sec.~\ref{sec:coincidence}.
These events constitute only 5\% of the sample of high-energy nuclear recoils, and the uncertainty in this correction is at most a fraction of this value.
Below 85\,k\eve , the shape of the nuclear recoil spectrum is a larger source of uncertainty.
We investigated several possible sources of spectral distortion, including the uncertainty in the energy calibration constant, the spectrum of residual electronic recoils subtracted from the high-energy nuclear recoil spectrum, 
and pileup.
We find the dominant uncertainty to be the assumption that the \iso{24}Na spectrum is an accurate model for the spectrum of gamma rays from the \ambe\ source below 140\,k\eve .
To estimate the impact of deviations in the spectral shape of the gamma-ray background, we reconstruct the low-energy nuclear-recoil spectrum after applying an exponential multiplicative correction to the \iso{24}Na data spectrum.
The correction was chosen to provide the best match between the simulated \iso{24}Na and \ambe spectra in Fig.~\ref{fig:gcomp}, with the main effect of increasing the subtracted gamma-ray background by at most 10\%.
This modification results in an increase in the defect-identification efficiency within the 1-$\sigma$ uncertainties in Fig.~\ref{fig:defficiency2}.

\section{Conclusion}
\label{sec:disc}

We have demonstrated for the first time nuclear recoil identification in a silicon CCD.
The experimental technique relies on the spatial correlation between the primary ionization event and the defect left behind by the atomic dislocation, later identified by thermal stimulation.
Since electronic recoils do not generate defects, this technique offers excellent discrimination between nuclear and electronic recoils down to nuclear recoil energies $<$10\,keV, competitive with other technologies for the direct detection of dark matter and CE$\nu$NS.
As presented in this article, the technique can be readily implemented in a CCD dark matter search (\emph{e.g.,} DAMIC-M~\cite{Privitera:2024tpq} and Oscura~\cite{Oscura:2022vmi}) to suppress electronic-recoil backgrounds by orders of magnitude and significantly increase sensitivity in the search for WIMPs with masses $>5$\,GeV\,$c^{-2}$.

This result is the first step to bring CCD experiments to the forefront in the search for nuclear recoils from weakly interacting particles.
Future work includes understanding the lifetime of defects at cold temperatures. If defects are sufficiently long lived such that the temperature cycle could be performed at most once a month, the deadtime in a dark matter search could be greatly reduced.

The presented strategy is limited at low nuclear-recoil energies by shot noise from leakage current in the warm images, which interferes with the identification of faint defects. Exploring in detail the temperature dependence of the signal from defects against the background from leakage current would help optimize the procedure.
Sensitivity could also be enhanced by stimulating the defects at lower temperatures, where leakage current is much smaller.
Possible techniques include Thermally Stimulated Current (TSC) analysis~\cite{Bruzzi:1992, *Pintilie:2002wx} and optical stimulation with near and short-wave infrared light~\cite{Scharf:2019oax}.

\begin{acknowledgments}
We are grateful to Alejandro Garc\'ia and Eric Smith at the Center for Experimental Nuclear Physics and Astrophysics (CENPA) for providing the radioactive sources for this study.
We thank Eric Dahl, Davide Franco, Ben Loer, Florian Reindl, Marco Rescigno, and Jingke Xu for insightful discussions on other detector technologies. 
We acknowledge financial support from the following agencies and organizations:
the U.S. Department of Energy Office of Science through the Dark Matter New Initiatives program;
the U.S. National Science Foundation through Grant No.\ NSF PHY-2110585 to the University of Washington and The University of Chicago;
Swiss National Science Foundation through Grant No. 200021\_153654 and via the Swiss Canton of Zurich;
IFCA through project PID2019-109829GB-I00 funded by MCIN/ AEI.
We thank the College of Arts and Sciences at the University of Washington for contributing the first CCDs to the DAMIC-M project.
The CCD development at Lawrence Berkeley National Laboratory MicroSystems Lab was supported in part by the Director, Office of Science, of the U.S. Department of Energy under Contract No. DE-AC02-05CH11231.
\end{acknowledgments}

\bibliography{biblio.bib}

\end{document}

%% file: preamble.tex
\usepackage{amsmath, textgreek, graphicx, grffile, float, enumerate, enumitem, color, natbib, hyperref, lipsum, soul}

\hypersetup{
	colorlinks = true,
	urlcolor = blue,
	citecolor = blue,
	linkcolor = blue,
	bookmarks = true,
	bookmarksopen = false
}

\usepackage[mathlines]{lineno}

\newcommand{\iso}[2]{$^{#1}$#2}                    \newcommand{\ambe}{$^{241}$Am$^{9}$Be }    
\newcommand{\eve}{eV$\mathrm{_{ee}}$}
\newcommand{\evr}{eV$\mathrm{_{nr}}$}
\newcommand{\sxy}{$\sigma_{xy}$}
\newcommand{\spix}{$\mathrm{\sigma_{pix}}$}


%% file: authors.tex
\author{K.J.\,McGuire} 
\affiliation{Center for Experimental Nuclear Physics and Astrophysics, University of Washington, Seattle, WA, United States}

\author{A.E.\,Chavarria}
\affiliation{Center for Experimental Nuclear Physics and Astrophysics, University of Washington, Seattle, WA, United States}

\author{N.\,Castell{o}-Mor}
\affiliation{Instituto de F\'{i}sica de Cantabria (IFCA), CSIC - Universidad de Cantabria, Santander, Spain}

\author{S.\,Lee}
\affiliation{Universit\"{a}t Z\"{u}rich Physik Institut, Z\"{u}rich, Switzerland}

\author{B.\,Kilminster}
\affiliation{Universit\"{a}t Z\"{u}rich Physik Institut, Z\"{u}rich, Switzerland}

\author{R.\,Vilar}
\affiliation{Instituto de F\'{i}sica de Cantabria (IFCA), CSIC - Universidad de Cantabria, Santander, Spain}

\author{A.\,Alvarez}
\affiliation{Center for Experimental Nuclear Physics and Astrophysics, University of Washington, Seattle, WA, United States}

\author{J.\,Jung}
\affiliation{Center for Experimental Nuclear Physics and Astrophysics, University of Washington, Seattle, WA, United States}

\author{J.\,Cuevas-Zepeda}
\affiliation{Kavli Institute for Cosmological Physics and The Enrico Fermi Institute, The University of Chicago, Chicago, IL, United States}

\author{C.\,De Dominicis}
\affiliation{Laboratoire de physique nucl\'{e}aire et des hautes \'{e}nergies (LPNHE), Sorbonne Universit\'{e}, Universit\'{e} Paris Cit\'{e}, CNRS/IN2P3, Paris, France}

\author{R.\,Ga\"{i}or}
\affiliation{Laboratoire de physique nucl\'{e}aire et des hautes \'{e}nergies (LPNHE), Sorbonne Universit\'{e}, Universit\'{e} Paris Cit\'{e}, CNRS/IN2P3, Paris, France}

\author{L.\,Iddir}
\affiliation{Laboratoire de physique nucl\'{e}aire et des hautes \'{e}nergies (LPNHE), Sorbonne Universit\'{e}, Universit\'{e} Paris Cit\'{e}, CNRS/IN2P3, Paris, France}

\author{A.\,Letessier-Selvon}
\affiliation{Laboratoire de physique nucl\'{e}aire et des hautes \'{e}nergies (LPNHE), Sorbonne Universit\'{e}, Universit\'{e} Paris Cit\'{e}, CNRS/IN2P3, Paris, France}

\author{H.\,Lin}
\affiliation{Center for Experimental Nuclear Physics and Astrophysics, University of Washington, Seattle, WA, United States}

\author{S.\,Munagavalasa}
\affiliation{Kavli Institute for Cosmological Physics and The Enrico Fermi Institute, The University of Chicago, Chicago, IL, United States}

\author{D.\,Norcini}
\affiliation{Kavli Institute for Cosmological Physics and The Enrico Fermi Institute, The University of Chicago, Chicago, IL, United States}

\author{S.\,Paul}
\affiliation{Kavli Institute for Cosmological Physics and The Enrico Fermi Institute, The University of Chicago, Chicago, IL, United States}

\author{P.\,Privitera}
\affiliation{Kavli Institute for Cosmological Physics and The Enrico Fermi Institute, The University of Chicago, Chicago, IL, United States}
\affiliation{Laboratoire de physique nucl\'{e}aire et des hautes \'{e}nergies (LPNHE), Sorbonne Universit\'{e}, Universit\'{e} Paris Cit\'{e}, CNRS/IN2P3, Paris, France}

\author{R.\,Smida}
\affiliation{Kavli Institute for Cosmological Physics and The Enrico Fermi Institute, The University of Chicago, Chicago, IL, United States}

\author{M.\,Traina}
\affiliation{Center for Experimental Nuclear Physics and Astrophysics, University of Washington, Seattle, WA, United States}

\author{R.\,Yajur}
\affiliation{Kavli Institute for Cosmological Physics and The Enrico Fermi Institute, The University of Chicago, Chicago, IL, United States}

\author{J-P.\,Zopounidis}
\affiliation{Laboratoire de physique nucl\'{e}aire et des hautes \'{e}nergies (LPNHE), Sorbonne Universit\'{e}, Universit\'{e} Paris Cit\'{e}, CNRS/IN2P3, Paris, France}

%% file: main.bbl
\begin{thebibliography}{46}%
\makeatletter
\providecommand \@ifxundefined [1]{%
 \@ifx{#1\undefined}
}%
\providecommand \@ifnum [1]{%
 \ifnum #1\expandafter \@firstoftwo
 \else \expandafter \@secondoftwo
 \fi
}%
\providecommand \@ifx [1]{%
 \ifx #1\expandafter \@firstoftwo
 \else \expandafter \@secondoftwo
 \fi
}%
\providecommand \natexlab [1]{#1}%
\providecommand \enquote  [1]{``#1''}%
\providecommand \bibnamefont  [1]{#1}%
\providecommand \bibfnamefont [1]{#1}%
\providecommand \citenamefont [1]{#1}%
\providecommand \href@noop [0]{\@secondoftwo}%
\providecommand \href [0]{\begingroup \@sanitize@url \@href}%
\providecommand \@href[1]{\@@startlink{#1}\@@href}%
\providecommand \@@href[1]{\endgroup#1\@@endlink}%
\providecommand \@sanitize@url [0]{\catcode `\\12\catcode `\$12\catcode
  `\&12\catcode `\#12\catcode `\^12\catcode `\_12\catcode `\%12\relax}%
\providecommand \@@startlink[1]{}%
\providecommand \@@endlink[0]{}%
\providecommand \url  [0]{\begingroup\@sanitize@url \@url }%
\providecommand \@url [1]{\endgroup\@href {#1}{\urlprefix }}%
\providecommand \urlprefix  [0]{URL }%
\providecommand \Eprint [0]{\href }%
\providecommand \doibase [0]{http://dx.doi.org/}%
\providecommand \selectlanguage [0]{\@gobble}%
\providecommand \bibinfo  [0]{\@secondoftwo}%
\providecommand \bibfield  [0]{\@secondoftwo}%
\providecommand \translation [1]{[#1]}%
\providecommand \BibitemOpen [0]{}%
\providecommand \bibitemStop [0]{}%
\providecommand \bibitemNoStop [0]{.\EOS\space}%
\providecommand \EOS [0]{\spacefactor3000\relax}%
\providecommand \BibitemShut  [1]{\csname bibitem#1\endcsname}%
\let\auto@bib@innerbib\@empty
\bibitem [{\citenamefont {Akerib}\ \emph {et~al.}(2022)\citenamefont {Akerib}
  \emph {et~al.}}]{DMDD2022}%
  \BibitemOpen
  \bibfield  {author} {\bibinfo {author} {\bibfnamefont {D.~S.}\ \bibnamefont
  {Akerib}} \emph {et~al.},\ }\bibfield  {title} {\enquote {\bibinfo {title}
  {{Dark Matter Direct Detection to the Neutrino Fog}},}\ }in\ \href@noop {}
  {\emph {\bibinfo {booktitle} {{Snowmass 2021}}}}\ (\bibinfo {year} {2022})\
  \Eprint {http://arxiv.org/abs/2203.08084} {arXiv:2203.08084 [hep-ex]}
  \BibitemShut {NoStop}%
\bibitem [{\citenamefont {Essig}\ \emph {et~al.}(2022)\citenamefont {Essig}
  \emph {et~al.}}]{Essig:2022dfa}%
  \BibitemOpen
  \bibfield  {author} {\bibinfo {author} {\bibfnamefont {Rouven}\ \bibnamefont
  {Essig}} \emph {et~al.},\ }\bibfield  {title} {\enquote {\bibinfo {title}
  {{Snowmass2021 Cosmic Frontier: The landscape of low-threshold dark matter
  direct detection in the next decade}},}\ }in\ \href@noop {} {\emph {\bibinfo
  {booktitle} {{Snowmass 2021}}}}\ (\bibinfo {year} {2022})\ \Eprint
  {http://arxiv.org/abs/2203.08297} {arXiv:2203.08297 [hep-ph]} \BibitemShut
  {NoStop}%
\bibitem [{\citenamefont {Abdullah}\ \emph {et~al.}(2022)\citenamefont
  {Abdullah} \emph {et~al.}}]{CEvNS2022}%
  \BibitemOpen
  \bibfield  {author} {\bibinfo {author} {\bibfnamefont {M.}~\bibnamefont
  {Abdullah}} \emph {et~al.},\ }\bibfield  {title} {\enquote {\bibinfo {title}
  {{Coherent elastic neutrino-nucleus scattering: Terrestrial and astrophysical
  applications}},}\ }in\ \href@noop {} {\emph {\bibinfo {booktitle} {{Snowmass
  2021}}}}\ (\bibinfo {year} {2022})\ \Eprint {http://arxiv.org/abs/2203.07361}
  {arXiv:2203.07361 [hep-ph]} \BibitemShut {NoStop}%
\bibitem [{\citenamefont {Armengaud}\ \emph {et~al.}(2016)\citenamefont
  {Armengaud} \emph {et~al.}}]{EDELWEISS:2016boq}%
  \BibitemOpen
  \bibfield  {author} {\bibinfo {author} {\bibfnamefont {E.}~\bibnamefont
  {Armengaud}} \emph {et~al.} (\bibinfo {collaboration} {EDELWEISS}),\
  }\bibfield  {title} {\enquote {\bibinfo {title} {{Constraints on low-mass
  WIMPs from the EDELWEISS-III dark matter search}},}\ }\href {\doibase
  10.1088/1475-7516/2016/05/019} {\bibfield  {journal} {\bibinfo  {journal}
  {JCAP}\ }\textbf {\bibinfo {volume} {05}},\ \bibinfo {pages} {019} (\bibinfo
  {year} {2016})},\ \Eprint {http://arxiv.org/abs/1603.05120} {arXiv:1603.05120
  [astro-ph.CO]} \BibitemShut {NoStop}%
\bibitem [{\citenamefont {Agnese}\ \emph {et~al.}(2014)\citenamefont {Agnese}
  \emph {et~al.}}]{SuperCDMS:2014cds}%
  \BibitemOpen
  \bibfield  {author} {\bibinfo {author} {\bibfnamefont {R.}~\bibnamefont
  {Agnese}} \emph {et~al.} (\bibinfo {collaboration} {SuperCDMS}),\ }\bibfield
  {title} {\enquote {\bibinfo {title} {{Search for Low-Mass Weakly Interacting
  Massive Particles with SuperCDMS}},}\ }\href {\doibase
  10.1103/PhysRevLett.112.241302} {\bibfield  {journal} {\bibinfo  {journal}
  {Phys. Rev. Lett.}\ }\textbf {\bibinfo {volume} {112}},\ \bibinfo {pages}
  {241302} (\bibinfo {year} {2014})},\ \Eprint {http://arxiv.org/abs/1402.7137}
  {arXiv:1402.7137 [hep-ex]} \BibitemShut {NoStop}%
\bibitem [{\citenamefont {Agnese}\ \emph {et~al.}(2018)\citenamefont {Agnese}
  \emph {et~al.}}]{SuperCDMS:2017mbc}%
  \BibitemOpen
  \bibfield  {author} {\bibinfo {author} {\bibfnamefont {R.}~\bibnamefont
  {Agnese}} \emph {et~al.} (\bibinfo {collaboration} {SuperCDMS}),\ }\bibfield
  {title} {\enquote {\bibinfo {title} {{Results from the Super Cryogenic Dark
  Matter Search Experiment at Soudan}},}\ }\href {\doibase
  10.1103/PhysRevLett.120.061802} {\bibfield  {journal} {\bibinfo  {journal}
  {Phys. Rev. Lett.}\ }\textbf {\bibinfo {volume} {120}},\ \bibinfo {pages}
  {061802} (\bibinfo {year} {2018})},\ \Eprint
  {http://arxiv.org/abs/1708.08869} {arXiv:1708.08869 [hep-ex]} \BibitemShut
  {NoStop}%
\bibitem [{\citenamefont {Abdelhameed}\ \emph {et~al.}(2019)\citenamefont
  {Abdelhameed} \emph {et~al.}}]{CRESST:2019jnq}%
  \BibitemOpen
  \bibfield  {author} {\bibinfo {author} {\bibfnamefont {A.~H.}\ \bibnamefont
  {Abdelhameed}} \emph {et~al.} (\bibinfo {collaboration} {CRESST}),\
  }\bibfield  {title} {\enquote {\bibinfo {title} {{First results from the
  CRESST-III low-mass dark matter program}},}\ }\href {\doibase
  10.1103/PhysRevD.100.102002} {\bibfield  {journal} {\bibinfo  {journal}
  {Phys. Rev. D}\ }\textbf {\bibinfo {volume} {100}},\ \bibinfo {pages}
  {102002} (\bibinfo {year} {2019})},\ \Eprint
  {http://arxiv.org/abs/1904.00498} {arXiv:1904.00498 [astro-ph.CO]}
  \BibitemShut {NoStop}%
\bibitem [{\citenamefont {Agnes}\ \emph {et~al.}(2018)\citenamefont {Agnes}
  \emph {et~al.}}]{DarkSide:2018kuk}%
  \BibitemOpen
  \bibfield  {author} {\bibinfo {author} {\bibfnamefont {P.}~\bibnamefont
  {Agnes}} \emph {et~al.} (\bibinfo {collaboration} {DarkSide}),\ }\bibfield
  {title} {\enquote {\bibinfo {title} {{DarkSide-50 532-day Dark Matter Search
  with Low-Radioactivity Argon}},}\ }\href {\doibase
  10.1103/PhysRevD.98.102006} {\bibfield  {journal} {\bibinfo  {journal} {Phys.
  Rev. D}\ }\textbf {\bibinfo {volume} {98}},\ \bibinfo {pages} {102006}
  (\bibinfo {year} {2018})},\ \Eprint {http://arxiv.org/abs/1802.07198}
  {arXiv:1802.07198 [astro-ph.CO]} \BibitemShut {NoStop}%
\bibitem [{\citenamefont {Aalbers}\ \emph {et~al.}(2023)\citenamefont {Aalbers}
  \emph {et~al.}}]{LUX-ZEPLIN:2022xrq}%
  \BibitemOpen
  \bibfield  {author} {\bibinfo {author} {\bibfnamefont {J.}~\bibnamefont
  {Aalbers}} \emph {et~al.} (\bibinfo {collaboration} {LUX-ZEPLIN}),\
  }\bibfield  {title} {\enquote {\bibinfo {title} {{First Dark Matter Search
  Results from the LUX-ZEPLIN (LZ) Experiment}},}\ }\href {\doibase
  10.1103/PhysRevLett.131.041002} {\bibfield  {journal} {\bibinfo  {journal}
  {Phys. Rev. Lett.}\ }\textbf {\bibinfo {volume} {131}},\ \bibinfo {pages}
  {041002} (\bibinfo {year} {2023})},\ \Eprint
  {http://arxiv.org/abs/2207.03764} {arXiv:2207.03764 [hep-ex]} \BibitemShut
  {NoStop}%
\bibitem [{\citenamefont {Holland}\ \emph {et~al.}(2003)\citenamefont
  {Holland}, \citenamefont {Groom}, \citenamefont {Palaio}, \citenamefont
  {Stover},\ and\ \citenamefont {Wei}}]{Holland:2003kiw}%
  \BibitemOpen
  \bibfield  {author} {\bibinfo {author} {\bibfnamefont {S.~E.}\ \bibnamefont
  {Holland}}, \bibinfo {author} {\bibfnamefont {D.~E.}\ \bibnamefont {Groom}},
  \bibinfo {author} {\bibfnamefont {N.~P.}\ \bibnamefont {Palaio}}, \bibinfo
  {author} {\bibfnamefont {R.~J.}\ \bibnamefont {Stover}}, \ and\ \bibinfo
  {author} {\bibfnamefont {Mingzhi}\ \bibnamefont {Wei}},\ }\bibfield  {title}
  {\enquote {\bibinfo {title} {{Fully depleted, back-illuminated charge-coupled
  devices fabricated on high-resistivity silicon}},}\ }\href {\doibase
  10.1109/TED.2002.806476} {\bibfield  {journal} {\bibinfo  {journal} {IEEE
  Trans. Electron. Dev.}\ }\textbf {\bibinfo {volume} {50}},\ \bibinfo {pages}
  {225--238} (\bibinfo {year} {2003})}\BibitemShut {NoStop}%
\bibitem [{\citenamefont {Tiffenberg}\ \emph {et~al.}(2017)\citenamefont
  {Tiffenberg}, \citenamefont {Sofo-Haro}, \citenamefont {Drlica-Wagner},
  \citenamefont {Essig}, \citenamefont {Guardincerri}, \citenamefont {Holland},
  \citenamefont {Volansky},\ and\ \citenamefont {Yu}}]{Tiffenberg:2017aac}%
  \BibitemOpen
  \bibfield  {author} {\bibinfo {author} {\bibfnamefont {Javier}\ \bibnamefont
  {Tiffenberg}}, \bibinfo {author} {\bibfnamefont {Miguel}\ \bibnamefont
  {Sofo-Haro}}, \bibinfo {author} {\bibfnamefont {Alex}\ \bibnamefont
  {Drlica-Wagner}}, \bibinfo {author} {\bibfnamefont {Rouven}\ \bibnamefont
  {Essig}}, \bibinfo {author} {\bibfnamefont {Yann}\ \bibnamefont
  {Guardincerri}}, \bibinfo {author} {\bibfnamefont {Steve}\ \bibnamefont
  {Holland}}, \bibinfo {author} {\bibfnamefont {Tomer}\ \bibnamefont
  {Volansky}}, \ and\ \bibinfo {author} {\bibfnamefont {Tien-Tien}\
  \bibnamefont {Yu}} (\bibinfo {collaboration} {SENSEI}),\ }\bibfield  {title}
  {\enquote {\bibinfo {title} {{Single-electron and single-photon sensitivity
  with a silicon Skipper CCD}},}\ }\href {\doibase
  10.1103/PhysRevLett.119.131802} {\bibfield  {journal} {\bibinfo  {journal}
  {Phys. Rev. Lett.}\ }\textbf {\bibinfo {volume} {119}},\ \bibinfo {pages}
  {131802} (\bibinfo {year} {2017})},\ \Eprint
  {http://arxiv.org/abs/1706.00028} {arXiv:1706.00028 [physics.ins-det]}
  \BibitemShut {NoStop}%
\bibitem [{\citenamefont {Aguilar-Arevalo}\ \emph {et~al.}(2020)\citenamefont
  {Aguilar-Arevalo} \emph {et~al.}}]{DAMIC:2020cut}%
  \BibitemOpen
  \bibfield  {author} {\bibinfo {author} {\bibfnamefont {A.}~\bibnamefont
  {Aguilar-Arevalo}} \emph {et~al.} (\bibinfo {collaboration} {DAMIC}),\
  }\bibfield  {title} {\enquote {\bibinfo {title} {{Results on low-mass weakly
  interacting massive particles from a 11 kg-day target exposure of DAMIC at
  SNOLAB}},}\ }\href {\doibase 10.1103/PhysRevLett.125.241803} {\bibfield
  {journal} {\bibinfo  {journal} {Phys. Rev. Lett.}\ }\textbf {\bibinfo
  {volume} {125}},\ \bibinfo {pages} {241803} (\bibinfo {year} {2020})},\
  \Eprint {http://arxiv.org/abs/2007.15622} {arXiv:2007.15622 [astro-ph.CO]}
  \BibitemShut {NoStop}%
\bibitem [{\citenamefont {Fernandez~Moroni}\ \emph {et~al.}(2015)\citenamefont
  {Fernandez~Moroni}, \citenamefont {Estrada}, \citenamefont {Paolini},
  \citenamefont {Cancelo}, \citenamefont {Tiffenberg},\ and\ \citenamefont
  {Molina}}]{CONNIE:2014qlq}%
  \BibitemOpen
  \bibfield  {author} {\bibinfo {author} {\bibfnamefont {Guillermo}\
  \bibnamefont {Fernandez~Moroni}}, \bibinfo {author} {\bibfnamefont {Juan}\
  \bibnamefont {Estrada}}, \bibinfo {author} {\bibfnamefont {Eduardo~E.}\
  \bibnamefont {Paolini}}, \bibinfo {author} {\bibfnamefont {Gustavo}\
  \bibnamefont {Cancelo}}, \bibinfo {author} {\bibfnamefont {Javier}\
  \bibnamefont {Tiffenberg}}, \ and\ \bibinfo {author} {\bibfnamefont {Jorge}\
  \bibnamefont {Molina}},\ }\bibfield  {title} {\enquote {\bibinfo {title}
  {{Charge Coupled Devices for detection of coherent neutrino-nucleus
  scattering}},}\ }\href {\doibase 10.1103/PhysRevD.91.072001} {\bibfield
  {journal} {\bibinfo  {journal} {Phys. Rev. D}\ }\textbf {\bibinfo {volume}
  {91}},\ \bibinfo {pages} {072001} (\bibinfo {year} {2015})},\ \Eprint
  {http://arxiv.org/abs/1405.5761} {arXiv:1405.5761 [physics.ins-det]}
  \BibitemShut {NoStop}%
\bibitem [{\citenamefont {Aguilar-Arevalo}\ \emph
  {et~al.}(2022{\natexlab{a}})\citenamefont {Aguilar-Arevalo} \emph
  {et~al.}}]{CONNIE:2021ggh}%
  \BibitemOpen
  \bibfield  {author} {\bibinfo {author} {\bibfnamefont {Alexis}\ \bibnamefont
  {Aguilar-Arevalo}} \emph {et~al.} (\bibinfo {collaboration} {CONNIE}),\
  }\bibfield  {title} {\enquote {\bibinfo {title} {{Search for coherent elastic
  neutrino-nucleus scattering at a nuclear reactor with CONNIE 2019 data}},}\
  }\href {\doibase 10.1007/JHEP05(2022)017} {\bibfield  {journal} {\bibinfo
  {journal} {JHEP}\ }\textbf {\bibinfo {volume} {05}},\ \bibinfo {pages} {017}
  (\bibinfo {year} {2022}{\natexlab{a}})},\ \Eprint
  {http://arxiv.org/abs/2110.13033} {arXiv:2110.13033 [hep-ex]} \BibitemShut
  {NoStop}%
\bibitem [{\citenamefont {Baxter}\ \emph {et~al.}(2020)\citenamefont {Baxter}
  \emph {et~al.}}]{Baxter:2019mcx}%
  \BibitemOpen
  \bibfield  {author} {\bibinfo {author} {\bibfnamefont {D.}~\bibnamefont
  {Baxter}} \emph {et~al.},\ }\bibfield  {title} {\enquote {\bibinfo {title}
  {{Coherent Elastic Neutrino-Nucleus Scattering at the European Spallation
  Source}},}\ }\href {\doibase 10.1007/JHEP02(2020)123} {\bibfield  {journal}
  {\bibinfo  {journal} {JHEP}\ }\textbf {\bibinfo {volume} {02}},\ \bibinfo
  {pages} {123} (\bibinfo {year} {2020})},\ \Eprint
  {http://arxiv.org/abs/1911.00762} {arXiv:1911.00762 [physics.ins-det]}
  \BibitemShut {NoStop}%
\bibitem [{\citenamefont {Lee}(2023{\natexlab{a}})}]{Lee:2022sxx}%
  \BibitemOpen
  \bibfield  {author} {\bibinfo {author} {\bibfnamefont {Steven~J.}\
  \bibnamefont {Lee}} (\bibinfo {collaboration} {DAMIC-M}),\ }\bibfield
  {title} {\enquote {\bibinfo {title} {{Analysis of radiation damage in silicon
  charge-coupled devices used for dark matter searches}},}\ }\href {\doibase
  10.21468/SciPostPhysProc.12.030} {\bibfield  {journal} {\bibinfo  {journal}
  {SciPost Phys. Proc.}\ }\textbf {\bibinfo {volume} {12}},\ \bibinfo {pages}
  {030} (\bibinfo {year} {2023}{\natexlab{a}})},\ \Eprint
  {http://arxiv.org/abs/2210.00469} {arXiv:2210.00469 [physics.ins-det]}
  \BibitemShut {NoStop}%
\bibitem [{\citenamefont {Lee}(2023{\natexlab{b}})}]{Steven-thesis}%
  \BibitemOpen
  \bibfield  {author} {\bibinfo {author} {\bibfnamefont {Steven~J.}\
  \bibnamefont {Lee}},\ }\emph {\bibinfo {title} {{Development of Technique to
  Use Lattice Defects in CCDs to Search for Dark Matter}}},\ \href@noop {}
  {Ph.D. thesis},\ \bibinfo  {school} {{University of Zurich}}, \bibinfo
  {address} {{Zurich, Switzerland}} (\bibinfo {year} {2023}{\natexlab{b}}),\
  \bibinfo {note} {{available at \url{https://www.zora.uzh.ch}}}\BibitemShut
  {NoStop}%
\bibitem [{\citenamefont {Nordlund}\ \emph {et~al.}(1998)\citenamefont
  {Nordlund}, \citenamefont {Ghaly}, \citenamefont {Averback}, \citenamefont
  {Caturla}, \citenamefont {Diaz de~la Rubia},\ and\ \citenamefont
  {Tarus}}]{Nordlund1998}%
  \BibitemOpen
  \bibfield  {author} {\bibinfo {author} {\bibfnamefont {K.}~\bibnamefont
  {Nordlund}}, \bibinfo {author} {\bibfnamefont {M.}~\bibnamefont {Ghaly}},
  \bibinfo {author} {\bibfnamefont {R.~S.}\ \bibnamefont {Averback}}, \bibinfo
  {author} {\bibfnamefont {M.}~\bibnamefont {Caturla}}, \bibinfo {author}
  {\bibfnamefont {T.}~\bibnamefont {Diaz de~la Rubia}}, \ and\ \bibinfo
  {author} {\bibfnamefont {J.}~\bibnamefont {Tarus}},\ }\bibfield  {title}
  {\enquote {\bibinfo {title} {Defect production in collision cascades in
  elemental semiconductors and fcc metals},}\ }\href {\doibase
  10.1103/PhysRevB.57.7556} {\bibfield  {journal} {\bibinfo  {journal} {Phys.
  Rev. B}\ }\textbf {\bibinfo {volume} {57}},\ \bibinfo {pages} {7556--7570}
  (\bibinfo {year} {1998})}\BibitemShut {NoStop}%
\bibitem [{\citenamefont {Moll}(1999)}]{Moll:1999kv}%
  \BibitemOpen
  \bibfield  {author} {\bibinfo {author} {\bibfnamefont {Michael}\ \bibnamefont
  {Moll}},\ }\emph {\bibinfo {title} {{Radiation damage in silicon particle
  detectors: Microscopic defects and macroscopic properties}}},\ \href@noop {}
  {Ph.D. thesis},\ \bibinfo  {school} {Hamburg U.} (\bibinfo {year} {1999}),\
  \bibinfo {note} {{available at
  \url{https://mmoll.web.cern.ch/thesis/}}}\BibitemShut {NoStop}%
\bibitem [{\citenamefont {Srour}\ \emph {et~al.}(2003)\citenamefont {Srour},
  \citenamefont {Marshall},\ and\ \citenamefont {Marshall}}]{Srour2003}%
  \BibitemOpen
  \bibfield  {author} {\bibinfo {author} {\bibfnamefont {J.R.}\ \bibnamefont
  {Srour}}, \bibinfo {author} {\bibfnamefont {C.J.}\ \bibnamefont {Marshall}},
  \ and\ \bibinfo {author} {\bibfnamefont {P.W.}\ \bibnamefont {Marshall}},\
  }\bibfield  {title} {\enquote {\bibinfo {title} {Review of displacement
  damage effects in silicon devices},}\ }\href {\doibase
  10.1109/TNS.2003.813197} {\bibfield  {journal} {\bibinfo  {journal} {IEEE
  Trans. Nucl. Sci.}\ }\textbf {\bibinfo {volume} {50}},\ \bibinfo {pages}
  {653--670} (\bibinfo {year} {2003})}\BibitemShut {NoStop}%
\bibitem [{\citenamefont {Sassi}\ \emph {et~al.}(2022)\citenamefont {Sassi},
  \citenamefont {Heikinheimo}, \citenamefont {Tuominen}, \citenamefont
  {Kuronen}, \citenamefont {Byggm\"astar}, \citenamefont {Nordlund},\ and\
  \citenamefont {Mirabolfathi}}]{Sassi:2022njl}%
  \BibitemOpen
  \bibfield  {author} {\bibinfo {author} {\bibfnamefont {Sebastian}\
  \bibnamefont {Sassi}}, \bibinfo {author} {\bibfnamefont {Matti}\ \bibnamefont
  {Heikinheimo}}, \bibinfo {author} {\bibfnamefont {Kimmo}\ \bibnamefont
  {Tuominen}}, \bibinfo {author} {\bibfnamefont {Antti}\ \bibnamefont
  {Kuronen}}, \bibinfo {author} {\bibfnamefont {Jesper}\ \bibnamefont
  {Byggm\"astar}}, \bibinfo {author} {\bibfnamefont {Kai}\ \bibnamefont
  {Nordlund}}, \ and\ \bibinfo {author} {\bibfnamefont {Nader}\ \bibnamefont
  {Mirabolfathi}},\ }\bibfield  {title} {\enquote {\bibinfo {title} {{Energy
  loss in low energy nuclear recoils in dark matter detector materials}},}\
  }\href {\doibase 10.1103/PhysRevD.106.063012} {\bibfield  {journal} {\bibinfo
   {journal} {Phys. Rev. D}\ }\textbf {\bibinfo {volume} {106}},\ \bibinfo
  {pages} {063012} (\bibinfo {year} {2022})},\ \Eprint
  {http://arxiv.org/abs/2206.06772} {arXiv:2206.06772 [hep-ph]} \BibitemShut
  {NoStop}%
\bibitem [{\citenamefont {Pintilie}\ \emph {et~al.}(2003)\citenamefont
  {Pintilie}, \citenamefont {Fretwurst}, \citenamefont {Lindstrom},\ and\
  \citenamefont {Stahl}}]{Pintilie:2002ge}%
  \BibitemOpen
  \bibfield  {author} {\bibinfo {author} {\bibfnamefont {I.}~\bibnamefont
  {Pintilie}}, \bibinfo {author} {\bibfnamefont {E.}~\bibnamefont {Fretwurst}},
  \bibinfo {author} {\bibfnamefont {G.}~\bibnamefont {Lindstrom}}, \ and\
  \bibinfo {author} {\bibfnamefont {J.}~\bibnamefont {Stahl}},\ }\bibfield
  {title} {\enquote {\bibinfo {title} {{Results on defects induced by Co-60
  gamma irradiation in standard and oxygen enriched silicon}},}\ }\href
  {\doibase 10.1016/j.nima.2003.08.079} {\bibfield  {journal} {\bibinfo
  {journal} {Nucl. Instrum. Meth. A}\ }\textbf {\bibinfo {volume} {514}},\
  \bibinfo {pages} {18--24} (\bibinfo {year} {2003})}\BibitemShut {NoStop}%
\bibitem [{\citenamefont {Fretwurst}\ \emph {et~al.}(2003)\citenamefont
  {Fretwurst}, \citenamefont {Lindstrom}, \citenamefont {Stahl}, \citenamefont
  {Pintilie}, \citenamefont {Li}, \citenamefont {Kierstead}, \citenamefont
  {Verbitskaya},\ and\ \citenamefont {Roder}}]{Fretwurst:2002gd}%
  \BibitemOpen
  \bibfield  {author} {\bibinfo {author} {\bibfnamefont {E.}~\bibnamefont
  {Fretwurst}}, \bibinfo {author} {\bibfnamefont {G.}~\bibnamefont
  {Lindstrom}}, \bibinfo {author} {\bibfnamefont {J.}~\bibnamefont {Stahl}},
  \bibinfo {author} {\bibfnamefont {I.}~\bibnamefont {Pintilie}}, \bibinfo
  {author} {\bibfnamefont {Z.}~\bibnamefont {Li}}, \bibinfo {author}
  {\bibfnamefont {J.}~\bibnamefont {Kierstead}}, \bibinfo {author}
  {\bibfnamefont {E.}~\bibnamefont {Verbitskaya}}, \ and\ \bibinfo {author}
  {\bibfnamefont {R.}~\bibnamefont {Roder}},\ }\bibfield  {title} {\enquote
  {\bibinfo {title} {{Bulk damage effects in standard and oxygen enriched
  silicon detectors induced by Co-60 gamma radiation}},}\ }\href {\doibase
  10.1016/j.nima.2003.08.077} {\bibfield  {journal} {\bibinfo  {journal} {Nucl.
  Instrum. Meth. A}\ }\textbf {\bibinfo {volume} {514}},\ \bibinfo {pages}
  {1--8} (\bibinfo {year} {2003})}\BibitemShut {NoStop}%
\bibitem [{\citenamefont {Arnquist}\ \emph {et~al.}(2023)\citenamefont
  {Arnquist} \emph {et~al.}}]{DAMIC-M:2023gxo}%
  \BibitemOpen
  \bibfield  {author} {\bibinfo {author} {\bibfnamefont {I.}~\bibnamefont
  {Arnquist}} \emph {et~al.} (\bibinfo {collaboration} {DAMIC-M}),\ }\bibfield
  {title} {\enquote {\bibinfo {title} {{First Constraints from DAMIC-M on
  Sub-GeV Dark-Matter Particles Interacting with Electrons}},}\ }\href
  {\doibase 10.1103/PhysRevLett.130.171003} {\bibfield  {journal} {\bibinfo
  {journal} {Phys. Rev. Lett.}\ }\textbf {\bibinfo {volume} {130}},\ \bibinfo
  {pages} {171003} (\bibinfo {year} {2023})},\ \Eprint
  {http://arxiv.org/abs/2302.02372} {arXiv:2302.02372 [hep-ex]} \BibitemShut
  {NoStop}%
\bibitem [{\citenamefont {Janesick}(2001)}]{janesick2001scientific}%
  \BibitemOpen
  \bibfield  {author} {\bibinfo {author} {\bibfnamefont {J.R.}\ \bibnamefont
  {Janesick}},\ }\href {https://books.google.com/books?id=rkgBkbDie7kC} {\emph
  {\bibinfo {title} {Scientific Charge-coupled Devices}}},\ Press Monograph
  Series\ (\bibinfo  {publisher} {Society of Photo Optical},\ \bibinfo {year}
  {2001})\BibitemShut {NoStop}%
\bibitem [{\citenamefont {Norcini}\ \emph {et~al.}(2022)\citenamefont {Norcini}
  \emph {et~al.}}]{DAMIC-M:2022xtp}%
  \BibitemOpen
  \bibfield  {author} {\bibinfo {author} {\bibfnamefont {D.}~\bibnamefont
  {Norcini}} \emph {et~al.} (\bibinfo {collaboration} {DAMIC-M}),\ }\bibfield
  {title} {\enquote {\bibinfo {title} {{Precision measurement of Compton
  scattering in silicon with a skipper CCD for dark matter detection}},}\
  }\href {\doibase 10.1103/PhysRevD.106.092001} {\bibfield  {journal} {\bibinfo
   {journal} {Phys. Rev. D}\ }\textbf {\bibinfo {volume} {106}},\ \bibinfo
  {pages} {092001} (\bibinfo {year} {2022})},\ \Eprint
  {http://arxiv.org/abs/2207.00809} {arXiv:2207.00809 [physics.ins-det]}
  \BibitemShut {NoStop}%
\bibitem [{\citenamefont {Ramanathan}\ and\ \citenamefont
  {Kurinsky}(2020)}]{Ramanathan:2020fwm}%
  \BibitemOpen
  \bibfield  {author} {\bibinfo {author} {\bibfnamefont {K.}~\bibnamefont
  {Ramanathan}}\ and\ \bibinfo {author} {\bibfnamefont {N.}~\bibnamefont
  {Kurinsky}},\ }\bibfield  {title} {\enquote {\bibinfo {title} {{Ionization
  yield in silicon for eV-scale electron-recoil processes}},}\ }\href {\doibase
  10.1103/PhysRevD.102.063026} {\bibfield  {journal} {\bibinfo  {journal}
  {Phys. Rev. D}\ }\textbf {\bibinfo {volume} {102}},\ \bibinfo {pages}
  {063026} (\bibinfo {year} {2020})},\ \Eprint
  {http://arxiv.org/abs/2004.10709} {arXiv:2004.10709 [astro-ph.IM]}
  \BibitemShut {NoStop}%
\bibitem [{\citenamefont {Sarkis}\ \emph {et~al.}(2023)\citenamefont {Sarkis},
  \citenamefont {Aguilar-Arevalo},\ and\ \citenamefont
  {D'Olivo}}]{Sarkis:2022pvc}%
  \BibitemOpen
  \bibfield  {author} {\bibinfo {author} {\bibfnamefont {Y.}~\bibnamefont
  {Sarkis}}, \bibinfo {author} {\bibfnamefont {A.}~\bibnamefont
  {Aguilar-Arevalo}}, \ and\ \bibinfo {author} {\bibfnamefont {J.~C.}\
  \bibnamefont {D'Olivo}},\ }\bibfield  {title} {\enquote {\bibinfo {title}
  {{Ionization efficiency for nuclear recoils in silicon from about 50~eV to 3
  MeV}},}\ }\href {\doibase 10.1103/PhysRevA.107.062811} {\bibfield  {journal}
  {\bibinfo  {journal} {Phys. Rev. A}\ }\textbf {\bibinfo {volume} {107}},\
  \bibinfo {pages} {062811} (\bibinfo {year} {2023})},\ \Eprint
  {http://arxiv.org/abs/2209.04503} {arXiv:2209.04503 [physics.atom-ph]}
  \BibitemShut {NoStop}%
\bibitem [{\citenamefont {Aguilar-Arevalo}\ \emph {et~al.}(2016)\citenamefont
  {Aguilar-Arevalo} \emph {et~al.}}]{DAMIC06KGD}%
  \BibitemOpen
  \bibfield  {author} {\bibinfo {author} {\bibfnamefont {A.}~\bibnamefont
  {Aguilar-Arevalo}} \emph {et~al.} (\bibinfo {collaboration} {DAMIC}),\
  }\bibfield  {title} {\enquote {\bibinfo {title} {{Search for low-mass WIMPs
  in a 0.6 kg day exposure of the DAMIC experiment at SNOLAB}},}\ }\href
  {\doibase 10.1103/PhysRevD.94.082006} {\bibfield  {journal} {\bibinfo
  {journal} {Phys. Rev. D}\ }\textbf {\bibinfo {volume} {94}},\ \bibinfo
  {pages} {082006} (\bibinfo {year} {2016})},\ \Eprint
  {http://arxiv.org/abs/1607.07410} {arXiv:1607.07410 [astro-ph.CO]}
  \BibitemShut {NoStop}%
\bibitem [{\citenamefont {Liu}\ \emph {et~al.}(2008)\citenamefont {Liu},
  \citenamefont {Chen}, \citenamefont {Zhu}, \citenamefont {Li},\ and\
  \citenamefont {Zhang}}]{liu:2007}%
  \BibitemOpen
  \bibfield  {author} {\bibinfo {author} {\bibfnamefont {Zhenzhou}\
  \bibnamefont {Liu}}, \bibinfo {author} {\bibfnamefont {Jinxiang}\
  \bibnamefont {Chen}}, \bibinfo {author} {\bibfnamefont {Pei}\ \bibnamefont
  {Zhu}}, \bibinfo {author} {\bibfnamefont {Yongming}\ \bibnamefont {Li}}, \
  and\ \bibinfo {author} {\bibfnamefont {Guohui}\ \bibnamefont {Zhang}},\
  }\bibfield  {title} {\enquote {\bibinfo {title} {{The 4.438MeV gamma to
  neutron ratio for the Am--Be neutron source}},}\ }\href {\doibase
  10.1016/j.apradiso.2007.04.007} {\bibfield  {journal} {\bibinfo  {journal}
  {Appl. Radiat. and Isot.}\ }\textbf {\bibinfo {volume} {65}},\ \bibinfo
  {pages} {1318--1321} (\bibinfo {year} {2008})}\BibitemShut {NoStop}%
\bibitem [{\citenamefont {Eriksen}\ \emph {et~al.}(2020)\citenamefont {Eriksen}
  \emph {et~al.}}]{Eriksen:2020isg}%
  \BibitemOpen
  \bibfield  {author} {\bibinfo {author} {\bibfnamefont {T.~K.}\ \bibnamefont
  {Eriksen}} \emph {et~al.},\ }\bibfield  {title} {\enquote {\bibinfo {title}
  {{Improved precision on the experimental E0 decay branching ratio of the
  Hoyle state}},}\ }\href {\doibase 10.1103/PhysRevC.102.024320} {\bibfield
  {journal} {\bibinfo  {journal} {Phys. Rev. C}\ }\textbf {\bibinfo {volume}
  {102}},\ \bibinfo {pages} {024320} (\bibinfo {year} {2020})},\ \Eprint
  {http://arxiv.org/abs/2007.15374} {arXiv:2007.15374 [nucl-ex]} \BibitemShut
  {NoStop}%
\bibitem [{\citenamefont {Basunia}(2006)}]{BASUNIA20062323}%
  \BibitemOpen
  \bibfield  {author} {\bibinfo {author} {\bibfnamefont {M.S.}\ \bibnamefont
  {Basunia}},\ }\bibfield  {title} {\enquote {\bibinfo {title} {Nuclear data
  sheets for a = 237},}\ }\href {\doibase
  https://doi.org/10.1016/j.nds.2006.07.001} {\bibfield  {journal} {\bibinfo
  {journal} {Nucl. Data Sheets}\ }\textbf {\bibinfo {volume} {107}},\ \bibinfo
  {pages} {2323--2422} (\bibinfo {year} {2006})}\BibitemShut {NoStop}%
\bibitem [{\citenamefont {Agostinelli}\ \emph {et~al.}(2003)\citenamefont
  {Agostinelli} \emph {et~al.}}]{GEANT4:2002zbu}%
  \BibitemOpen
  \bibfield  {author} {\bibinfo {author} {\bibfnamefont {S.}~\bibnamefont
  {Agostinelli}} \emph {et~al.} (\bibinfo {collaboration} {GEANT4}),\
  }\bibfield  {title} {\enquote {\bibinfo {title} {{GEANT4--a simulation
  toolkit}},}\ }\href {\doibase 10.1016/S0168-9002(03)01368-8} {\bibfield
  {journal} {\bibinfo  {journal} {Nucl. Instrum. Meth. A}\ }\textbf {\bibinfo
  {volume} {506}},\ \bibinfo {pages} {250--303} (\bibinfo {year}
  {2003})}\BibitemShut {NoStop}%
\bibitem [{\citenamefont {Kluge}\ and\ \citenamefont {Weise}(1982)}]{AmBeSpec}%
  \BibitemOpen
  \bibfield  {author} {\bibinfo {author} {\bibfnamefont {H.}~\bibnamefont
  {Kluge}}\ and\ \bibinfo {author} {\bibfnamefont {K.}~\bibnamefont {Weise}},\
  }\bibfield  {title} {\enquote {\bibinfo {title} {{The Neutron Energy Spectrum
  of a 241Am-Be(Alpha,n) Source and Resulting Mean Fluence to Dose Equivalent
  Conversion Factors}},}\ }\href {\doibase 10.1093/oxfordjournals.rpd.a080571}
  {\bibfield  {journal} {\bibinfo  {journal} {Radiat. Prot. Dosim.}\ }\textbf
  {\bibinfo {volume} {2}},\ \bibinfo {pages} {85--93} (\bibinfo {year}
  {1982})}\BibitemShut {NoStop}%
\bibitem [{\citenamefont {Duke}\ \emph {et~al.}(2016)\citenamefont {Duke},
  \citenamefont {Hallin}, \citenamefont {Krauss}, \citenamefont {Mekarski},\
  and\ \citenamefont {Sibley}}]{Duke:2015wga}%
  \BibitemOpen
  \bibfield  {author} {\bibinfo {author} {\bibfnamefont {M.~J.~M.}\
  \bibnamefont {Duke}}, \bibinfo {author} {\bibfnamefont {A.~L.}\ \bibnamefont
  {Hallin}}, \bibinfo {author} {\bibfnamefont {C.~B.}\ \bibnamefont {Krauss}},
  \bibinfo {author} {\bibfnamefont {P.}~\bibnamefont {Mekarski}}, \ and\
  \bibinfo {author} {\bibfnamefont {L.}~\bibnamefont {Sibley}},\ }\bibfield
  {title} {\enquote {\bibinfo {title} {{A precise method to determine the
  activity of a weak neutron source using a germanium detector}},}\ }\href
  {\doibase 10.1016/j.apradiso.2016.06.032} {\bibfield  {journal} {\bibinfo
  {journal} {Appl. Radiat. Isot.}\ }\textbf {\bibinfo {volume} {116}},\
  \bibinfo {pages} {51--56} (\bibinfo {year} {2016})},\ \Eprint
  {http://arxiv.org/abs/1506.05417} {arXiv:1506.05417 [physics.ins-det]}
  \BibitemShut {NoStop}%
\bibitem [{\citenamefont {Basunia}\ and\ \citenamefont
  {Chakraborty}(2022)}]{Basunia:2022zmp}%
  \BibitemOpen
  \bibfield  {author} {\bibinfo {author} {\bibfnamefont {M.~Shamsuzzoha}\
  \bibnamefont {Basunia}}\ and\ \bibinfo {author} {\bibfnamefont {Anagha}\
  \bibnamefont {Chakraborty}},\ }\bibfield  {title} {\enquote {\bibinfo {title}
  {{Nuclear Data Sheets for A=24}},}\ }\href {\doibase
  10.1016/j.nds.2022.11.002} {\bibfield  {journal} {\bibinfo  {journal} {Nucl.
  Data Sheets}\ }\textbf {\bibinfo {volume} {186}},\ \bibinfo {pages} {3--262}
  (\bibinfo {year} {2022})}\BibitemShut {NoStop}%
\bibitem [{\citenamefont {Agnese}\ \emph {et~al.}(2013)\citenamefont {Agnese}
  \emph {et~al.}}]{SuperCDMSSoudan:2013ukv}%
  \BibitemOpen
  \bibfield  {author} {\bibinfo {author} {\bibfnamefont {R.}~\bibnamefont
  {Agnese}} \emph {et~al.} (\bibinfo {collaboration} {SuperCDMSSoudan}),\
  }\bibfield  {title} {\enquote {\bibinfo {title} {{Demonstration of Surface
  Electron Rejection with Interleaved Germanium Detectors for Dark Matter
  Searches}},}\ }\href {\doibase 10.1063/1.4826093} {\bibfield  {journal}
  {\bibinfo  {journal} {Appl. Phys. Lett.}\ }\textbf {\bibinfo {volume}
  {103}},\ \bibinfo {pages} {164105} (\bibinfo {year} {2013})},\ \Eprint
  {http://arxiv.org/abs/1305.2405} {arXiv:1305.2405 [physics.ins-det]}
  \BibitemShut {NoStop}%
\bibitem [{\citenamefont {Strauss}\ \emph {et~al.}(2014)\citenamefont {Strauss}
  \emph {et~al.}}]{Strauss:2014zia}%
  \BibitemOpen
  \bibfield  {author} {\bibinfo {author} {\bibfnamefont {R.}~\bibnamefont
  {Strauss}} \emph {et~al.},\ }\bibfield  {title} {\enquote {\bibinfo {title}
  {{Energy-dependent light quenching in CaWO$_4$ crystals at mK
  temperatures}},}\ }\href {\doibase 10.1140/epjc/s10052-014-2957-5} {\bibfield
   {journal} {\bibinfo  {journal} {Eur. Phys. J. C}\ }\textbf {\bibinfo
  {volume} {74}},\ \bibinfo {pages} {2957} (\bibinfo {year} {2014})},\ \Eprint
  {http://arxiv.org/abs/1401.3332} {arXiv:1401.3332 [astro-ph.IM]} \BibitemShut
  {NoStop}%
\bibitem [{\citenamefont {Armengaud}\ \emph {et~al.}(2017)\citenamefont
  {Armengaud} \emph {et~al.}}]{EDELWEISS:2017lvq}%
  \BibitemOpen
  \bibfield  {author} {\bibinfo {author} {\bibfnamefont {E.}~\bibnamefont
  {Armengaud}} \emph {et~al.} (\bibinfo {collaboration} {EDELWEISS}),\
  }\bibfield  {title} {\enquote {\bibinfo {title} {{Performance of the
  EDELWEISS-III experiment for direct dark matter searches}},}\ }\href
  {\doibase 10.1088/1748-0221/12/08/P08010} {\bibfield  {journal} {\bibinfo
  {journal} {JINST}\ }\textbf {\bibinfo {volume} {12}},\ \bibinfo {pages}
  {P08010} (\bibinfo {year} {2017})},\ \Eprint
  {http://arxiv.org/abs/1706.01070} {arXiv:1706.01070 [physics.ins-det]}
  \BibitemShut {NoStop}%
\bibitem [{\citenamefont {Akerib}\ \emph {et~al.}(2020)\citenamefont {Akerib}
  \emph {et~al.}}]{LUX:2020car}%
  \BibitemOpen
  \bibfield  {author} {\bibinfo {author} {\bibfnamefont {D.~S.}\ \bibnamefont
  {Akerib}} \emph {et~al.} (\bibinfo {collaboration} {LUX}),\ }\bibfield
  {title} {\enquote {\bibinfo {title} {{Discrimination of electronic recoils
  from nuclear recoils in two-phase xenon time projection chambers}},}\ }\href
  {\doibase 10.1103/PhysRevD.102.112002} {\bibfield  {journal} {\bibinfo
  {journal} {Phys. Rev. D}\ }\textbf {\bibinfo {volume} {102}},\ \bibinfo
  {pages} {112002} (\bibinfo {year} {2020})},\ \Eprint
  {http://arxiv.org/abs/2004.06304} {arXiv:2004.06304 [physics.ins-det]}
  \BibitemShut {NoStop}%
\bibitem [{\citenamefont {Ali}\ \emph {et~al.}(2022)\citenamefont {Ali} \emph
  {et~al.}}]{PICO:2022nyi}%
  \BibitemOpen
  \bibfield  {author} {\bibinfo {author} {\bibfnamefont {B.}~\bibnamefont
  {Ali}} \emph {et~al.} (\bibinfo {collaboration} {PICO}),\ }\bibfield  {title}
  {\enquote {\bibinfo {title} {{Determining the bubble nucleation efficiency of
  low-energy nuclear recoils in superheated C$_3$F$_8$ dark matter
  detectors}},}\ }\href {\doibase 10.1103/PhysRevD.106.122003} {\bibfield
  {journal} {\bibinfo  {journal} {Phys. Rev. D}\ }\textbf {\bibinfo {volume}
  {106}},\ \bibinfo {pages} {122003} (\bibinfo {year} {2022})},\ \Eprint
  {http://arxiv.org/abs/2205.05771} {arXiv:2205.05771 [physics.ins-det]}
  \BibitemShut {NoStop}%
\bibitem [{\citenamefont {Privitera}(2024)}]{Privitera:2024tpq}%
  \BibitemOpen
  \bibfield  {author} {\bibinfo {author} {\bibfnamefont {Paolo}\ \bibnamefont
  {Privitera}} (\bibinfo {collaboration} {DAMIC-M}),\ }\bibfield  {title}
  {\enquote {\bibinfo {title} {{The DAMIC-M experiment: status and first
  results}},}\ }\href {\doibase 10.22323/1.441.0066} {\bibfield  {journal}
  {\bibinfo  {journal} {PoS}\ }\textbf {\bibinfo {volume} {TAUP2023}},\
  \bibinfo {pages} {066} (\bibinfo {year} {2024})}\BibitemShut {NoStop}%
\bibitem [{\citenamefont {Aguilar-Arevalo}\ \emph
  {et~al.}(2022{\natexlab{b}})\citenamefont {Aguilar-Arevalo} \emph
  {et~al.}}]{Oscura:2022vmi}%
  \BibitemOpen
  \bibfield  {author} {\bibinfo {author} {\bibfnamefont {Alexis}\ \bibnamefont
  {Aguilar-Arevalo}} \emph {et~al.} (\bibinfo {collaboration} {Oscura}),\
  }\bibfield  {title} {\enquote {\bibinfo {title} {{The Oscura Experiment}},}\
  }\href@noop {} {\  (\bibinfo {year} {2022}{\natexlab{b}})},\ \Eprint
  {http://arxiv.org/abs/2202.10518} {arXiv:2202.10518 [astro-ph.IM]}
  \BibitemShut {NoStop}%
\bibitem [{\citenamefont {Bruzzi}\ \emph {et~al.}(1992)\citenamefont {Bruzzi},
  \citenamefont {Borchi},\ and\ \citenamefont {Baldini}}]{Bruzzi:1992}%
  \BibitemOpen
  \bibfield  {author} {\bibinfo {author} {\bibfnamefont {M.}~\bibnamefont
  {Bruzzi}}, \bibinfo {author} {\bibfnamefont {E.}~\bibnamefont {Borchi}}, \
  and\ \bibinfo {author} {\bibfnamefont {A.}~\bibnamefont {Baldini}},\
  }\bibfield  {title} {\enquote {\bibinfo {title} {{Using thermally stimulated
  currents to visualize defect clusters in neutron‐irradiated silicon}},}\
  }\href {\doibase 10.1063/1.352253} {\bibfield  {journal} {\bibinfo  {journal}
  {J. Appl. Phys.}\ }\textbf {\bibinfo {volume} {72}},\ \bibinfo {pages} {4007}
  (\bibinfo {year} {1992})}\BibitemShut {NoStop}%
\bibitem [{\citenamefont {Pintilie}\ \emph {et~al.}(2002)\citenamefont
  {Pintilie}, \citenamefont {Tivarus}, \citenamefont {Pintilie}, \citenamefont
  {Moll}, \citenamefont {Fretwurst},\ and\ \citenamefont
  {Lindstrom}}]{Pintilie:2002wx}%
  \BibitemOpen
  \bibfield  {author} {\bibinfo {author} {\bibfnamefont {I.}~\bibnamefont
  {Pintilie}}, \bibinfo {author} {\bibfnamefont {C.}~\bibnamefont {Tivarus}},
  \bibinfo {author} {\bibfnamefont {L.}~\bibnamefont {Pintilie}}, \bibinfo
  {author} {\bibfnamefont {M.}~\bibnamefont {Moll}}, \bibinfo {author}
  {\bibfnamefont {E.}~\bibnamefont {Fretwurst}}, \ and\ \bibinfo {author}
  {\bibfnamefont {G.}~\bibnamefont {Lindstrom}},\ }\bibfield  {title} {\enquote
  {\bibinfo {title} {{Thermally stimulated current method applied to highly
  irradiated silicon diodes}},}\ }\href {\doibase
  10.1016/S0168-9002(01)01654-0} {\bibfield  {journal} {\bibinfo  {journal}
  {Nucl. Instrum. Meth. A}\ }\textbf {\bibinfo {volume} {476}},\ \bibinfo
  {pages} {652--657} (\bibinfo {year} {2002})}\BibitemShut {NoStop}%
\bibitem [{\citenamefont {Scharf}\ \emph {et~al.}(2020)\citenamefont {Scharf},
  \citenamefont {Feindt},\ and\ \citenamefont {Klanner}}]{Scharf:2019oax}%
  \BibitemOpen
  \bibfield  {author} {\bibinfo {author} {\bibfnamefont {C.}~\bibnamefont
  {Scharf}}, \bibinfo {author} {\bibfnamefont {F.}~\bibnamefont {Feindt}}, \
  and\ \bibinfo {author} {\bibfnamefont {R.}~\bibnamefont {Klanner}},\
  }\bibfield  {title} {\enquote {\bibinfo {title} {{Influence of radiation
  damage on the absorption of near-infrared light in silicon}},}\ }\href
  {\doibase 10.1016/j.nima.2020.163955} {\bibfield  {journal} {\bibinfo
  {journal} {Nucl. Instrum. Meth. A}\ }\textbf {\bibinfo {volume} {968}},\
  \bibinfo {pages} {163955} (\bibinfo {year} {2020})},\ \Eprint
  {http://arxiv.org/abs/1905.03874} {arXiv:1905.03874 [physics.ins-det]}
  \BibitemShut {NoStop}%
\end{thebibliography}%
